\documentclass[10pt]{article}
\usepackage[a4paper, margin=1in]{geometry}
\usepackage{authblk}
\usepackage{hyperref}
\usepackage{graphicx}
\usepackage{subcaption}
\usepackage{amssymb}
\usepackage{amsmath}
\usepackage{mathtools}
\usepackage{multirow}
\usepackage{booktabs}
\usepackage{threeparttable}
\usepackage{booktabs}
\usepackage{bm}
\usepackage{tikz}
\usepackage{rotating}
\usepackage{placeins}
\usepackage{pdflscape}
\usepackage{adjustbox}
\usepackage[style=apa,backend=biber,sorting=nyt]{biblatex}
\DeclareLanguageMapping{american}{american-apa}
\newcommand{\ubar}[1]{\underline{#1}}
\begin{document}

\title{Nonlinear Modal Interval Regression for Bivariate Data Analysis}

\author[1]{Sai Yao}
\author[1]{Yuko Araki\thanks{Corresponding author. Email: \href{mailto:yaraki@tohoku.ac.jp}{yaraki@tohoku.ac.jp}}}
\author[2]{Osuke Iwata}
\affil[1]{Graduate School of Information Sciences, Tohoku University, Sendai, Japan}
\affil[2]{Graduate School of Medical Sciences, Nagoya City University, Nagoya, Japan}
\date{} 

\maketitle

\begin{abstract}
The dispersion of real data is particularly important to understand the variability of a given distribution. In addition to the central tendency, variability is of considerable interest in a wide variety of fields such as life sciences, meteorology, and economics.
The modal interval (MI) describes the dispersion or spread of distribution and represents the most concentrated interval of a univariate unimodal distribution.
In this study, we propose a nonlinear modal interval regression (MIR) method to smoothly estimate a conditional MI to provide a robust description of how the dispersion of a data distribution varies with the covariate.
First, we use kernel density estimation (KDE) to estimate the quantile levels corresponding to the conditional MI bounds, which serve as input to the quantile loss function.
Second, we fit upper and lower bound functions using the quantile loss with smoothing splines.
The results of numerical experiments demonstrate that the reformulated MIR achieved higher accuracy and stability than both the conventional MIR and the KDE methods.
To evaluate the effectiveness of the proposed approach, we applied the method to neonatal hormone data and identified notable rhythms in cortisol and melatonin levels during the first ten days after birth.

\end{abstract}


\newpage
\section{Introduction}
\begin{figure}[t]
  \centering
  \begin{subfigure}{0.48\textwidth}
    \centering
    \includegraphics[width=\textwidth]{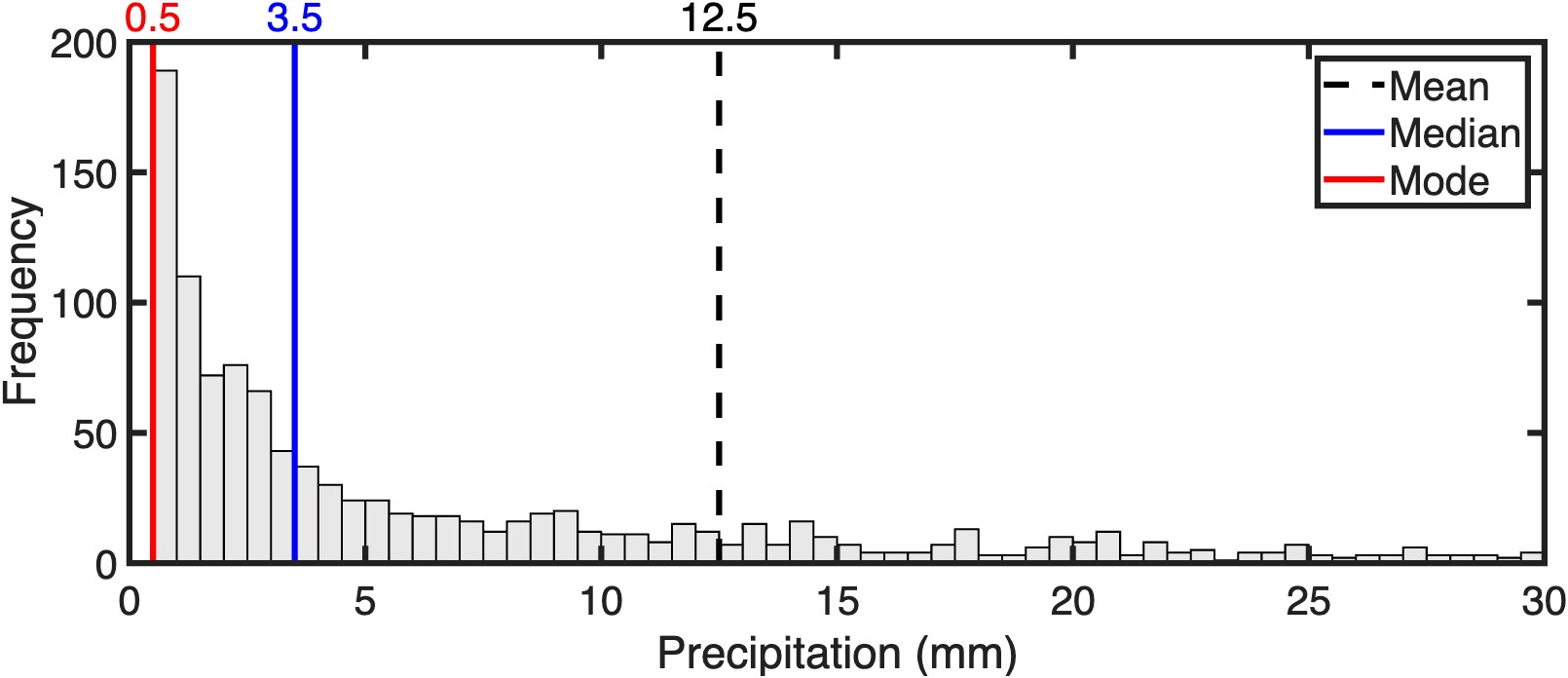}
    \caption{}
    \label{fig:rain_a}
  \end{subfigure}
    \hfill
  \begin{subfigure}{0.48\textwidth}
    \centering
    \includegraphics[width=\textwidth]{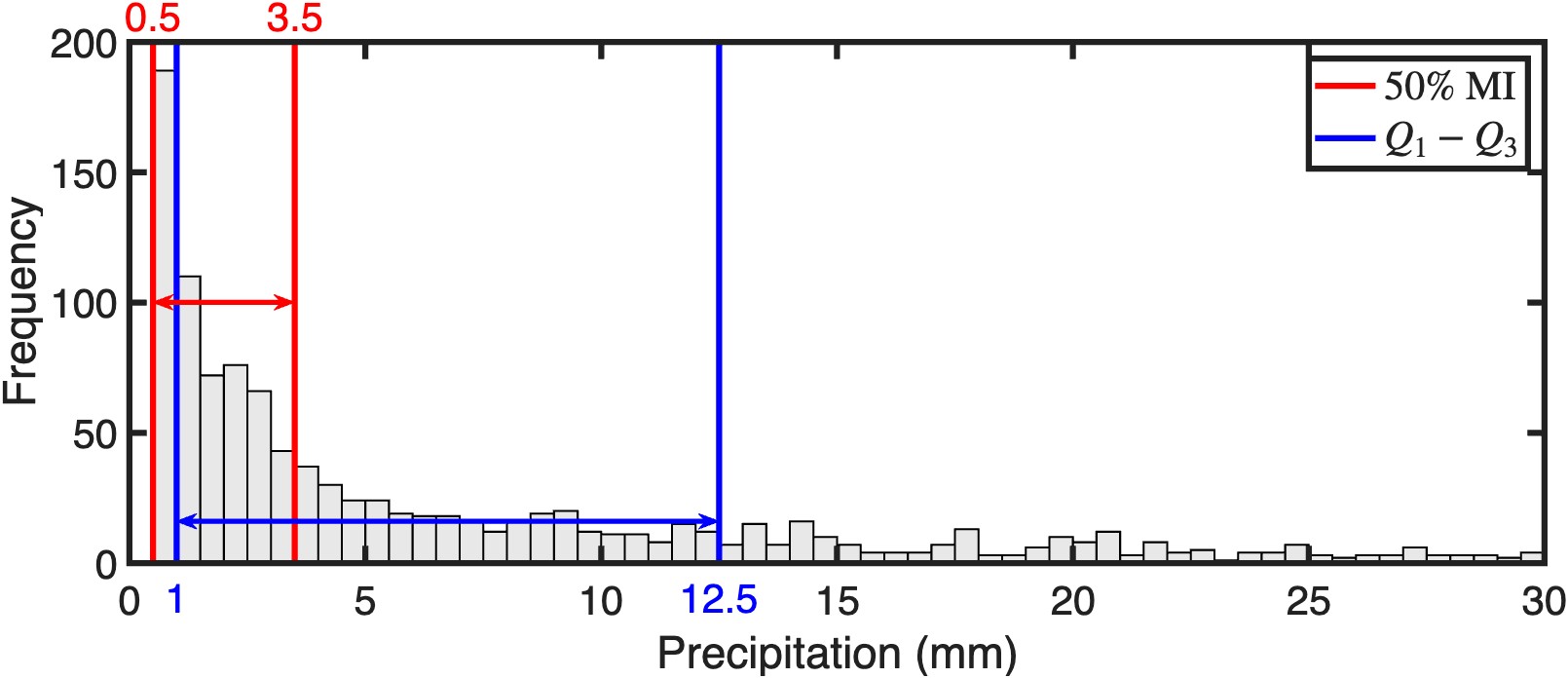}
    \caption{}
    \label{fig:rain_b}
  \end{subfigure}
  \caption{Daily precipitation ($\geq$0.1 mm) in Sendai, Japan (July 2014--July 2024). 
  (a) The blue solid line represents the mean, the orange dashed line represents the median, and the red dotted line represents the mode. 
  (b) The blue interval represents the range between the first ($Q_1$) and third ($Q_3$) quartiles, the widths of which correspond to the IQR, and the red interval represents the 50\% MI.}
  \label{fig:precipitation}
\end{figure}

Most works on bivariate regression analysis have focused on mean regression, which estimates the conditional mean of $Y$ given $X=x$ \parencite{Hastie2009,Khan2025}.
However, the mean is sensitive to outliers, which can make robust alternatives such as median regression \parencite{Bassett1978,maronna2019} and modal regression \parencite{Lee1989,Chen2016,Liu2024} more preferable in certain cases.
Although the mean, median, and mode coincide in symmetric unimodal distributions, they diverge in asymmetric distributions, and the mode offers greater robustness to outliers.
In many scientific fields, data often exhibit skewed or heavy-tailed distributions, such as with biological measurements, income, or precipitation.
As an example, Figure~\ref{fig:rain_a} shows daily precipitation (\(\geq 0.1\) mm) in Sendai, Japan, from July 2014 to July 2024. It may be observed that the mode remains stable despite the presence of outliers.

However, in many practical applications with real data, the dispersion of a distribution can be as important as the central tendency because measures of variability play an essential role in a wide variety of fields, from life science to economics and meteorology.
Measures such as the standard deviation (SD), interquartile range (IQR), and modal interval (MI; see Definition~1) are commonly used to assess the dispersion of a distribution.
In particular, the IQR is widely used for its robustness to outliers, whereas the MI is particularly effective for unimodal distributions because it directly captures the data concentration interval.
Figure~\ref{fig:rain_b} compares the interval between the first and third quartiles and the 50\% MI.
Although both cover $50\%$ of the data, the MI is much narrower, which highlights its superior ability to represent the data concentration interval more effectively.

In regression analysis, the conditional IQR is derived from quantile regression \parencite{Koenker1978,Koenker2005,Koenker2017}, which estimates the conditional quantile functions at the 25th and 75th percentiles.
While quantile regression captures distributional extremes and modal regression describes the most probable value, modal interval regression (MIR) provides a bridge between them by offering a compact view of where the data are most concentrated.
However, research on MIR remains limited, and recent studies on regression have increasingly shifted toward distributional regression \parencite{Chen2023,Wang2025} and Bayesian frameworks \parencite{Tanabe2022,Liu2024}.

The key challenge in MIR is that traditional regression methods estimate a single value, whereas MIR techniques estimate an entire interval.
Furthermore, the MI is estimated directly from the data without assuming any specific distribution.
Many methods simplify the problem by assuming normality because of the complexity of estimating the conditional MI without making assumptions about the distribution.
This reduction allows the task of estimating an interval to be framed as that of computing an equal-tailed interval with a specified level of coverage, which makes it considerably easier to estimate the data concentration interval.

Several approaches have been developed under the assumption of normality, including quantile regression, neural networks (NNs), and bootstrap-based methods.
For low-dimensional data, quantile regression can be applied to estimate the upper and lower bounds of the interval separately.
In contrast NN-based methods have been widely applied for higher-dimensional data  \parencite{Heskes1996,DeVeaux1998,Khosravi2010,Khosravi2011,Hwang1997,Shrestha2006}.
Bootstrap-based approaches have also been explored by \textcite{Kim2010,Beyaztas2020,Hwang2022}.
Although these approaches aim to construct prediction intervals rather than MI, they are often regarded as simplified cases of MI estimation under normality.
For normal data distributions, the estimated interval coincides with the MI.
However, in asymmetric unimodal distributions, the interval often becomes too wide to accurately represent the region in which the data are most concentrated.

Kernel density estimation (KDE) is commonly applied as a nonparametric method to estimate and visualize the conditional MI for bivariate data without assuming a specific distribution \parencite{DeGooijer2000,Polonik2000}.
KDE directly estimates the conditional distribution from local data to extract MI.
This approach is particularly useful when the data distribution is unknown because it does not rely on parametric assumptions.
Some methods have been extended to multivariate analysis, such as the random forest models proposed by \textcite{Zhu2019,Roy2020}.
However, these methods are  not suitable for bivariate analysis and visualization due to the inherent difficulty in interpreting their results in low-dimensional settings.
Although KDE is flexible and makes no strict assumptions, it involves some notable drawbacks such as limited estimation accuracy and a tendency to produce results that can be rough and unnatural. Given these limitations, KDE is considered less ideal for visualization.

To address these issues, \textcite{Yao2023} proposed a nonlinear MIR framework.
Their method was developed to estimate and visualize the conditional MI for bivariate data without assuming a specific parametric distribution.
By formulating the problem within a regression framework, MIR enables a smooth functional representation that enhances interpretability and helps to visualize the data structure.

The framework simplifies the complex task of estimating an entire conditional MI into two simpler subproblems in which its upper and lower bounds are estimated separately.
The method first divides the data into bins along the independent variable \(X\).
Within each bin, the MI of the dependent variable \(Y\) is estimated from its empirical distribution and then converted into the corresponding quantile levels.
Two polynomial functions are then fitted using the quantile loss to estimate the upper and lower bounds, which are smoothly connected across bins to ensure global continuity.
This bin-based MIR framework can be used to visualize the concentration interval of bivariate data without relying on parametric assumptions and achieves higher accuracy than existing methods.

Nonetheless, bin-based MIR frameworks also involve several important limitations.
First, its accuracy strongly depends on the binning strategy.
When the quantity of available data is limited, selecting appropriate bins can be difficult.
Bins that are too wide blur local structures, whereas bins that are too narrow produce unstable estimates. Both issues can reduce accuracy considerably.
Second, MI is estimated from the empirical distribution of \(Y\) within each bin instead of the conditional distribution of \(Y\) given \(X=x\).
This mismatch with the definition of conditional MI reduces the accuracy of the estimation, particularly when the conditional distribution varies sharply across \(X\).
Finally, the framework lacks a principled approach to tune the parameter selection, which makes it difficult to choose parameters objectively in applications with real data.

To resolve these issues, we reformulate the MIR framework by revising how the quantile level parameters in the quantile loss are obtained for fitting the upper and lower bounds.
Our contribution lies in this reformulation of the MIR framework.
Specifically, we make two methodological contributions.

First, we use KDE to estimate the quantile levels corresponding to the conditional MI bounds at each observed covariate value.
These estimated quantile levels are then used as the quantile level parameters in the quantile loss, and the upper and lower bounds are fitted using smoothing splines.
This approach avoids bin selection, is more consistent with the definition of the conditional MI, and adapts well to local distributional changes.
In addition, it separates spline knot selection from the estimation of the quantile level parameters, making the method more flexible in practice.

Second, we introduce a principled strategy to select parameters to enhance accuracy, stability, and reproducibility in practical applications.
Specifically, we use a plug-in method to select the KDE bandwidth automatically, and we choose the spline smoothing parameter via a modified criterion that better reflects the objective of MI estimation.

Numerical experiments show that the proposed method improves both accuracy and stability compared with KDE and bin-based MIR methods, particularly for complex distributions and when the sample size is limited.
We further provide an empirical assessment of practical behavior by examining
(i) how estimation accuracy and stability change as the sample size increases,
(ii) how performance varies with the smoothing parameter,
and (iii) how the computation time scales with the sample size.

When applied to neonatal hormone data, the proposed MIR method exhibited notable practical utility in analyzing complex biological rhythms.
Hormone levels in newborns exhibit substantial variability between individuals as well as long-tailed distributions, which make it difficult for conventional regression approaches to extract meaningful patterns.
Describing neonatal hormone rhythms with a single representative value can be challenging because outliers tend to affect the estimated trends, and the data contain large differences between individuals.
The proposed MIR method is robust to outliers and captures the most concentrated interval of the data effectively.
By analyzing the 50\% MI, MIR clearly shows how the most concentrated interval of neonatal hormone levels varies over time.
Thus, our results providing a clearer and more reliable depiction of the typical pre-circadian pattern.

The remainder of this study is organized as follows.
Section~2 presents the proposed nonlinear MIR, which addresses the limitations of bin-based MIR methods by using KDE to estimate the quantile levels corresponding to the conditional MI bounds.
In Section~3, we describe a numerical validation of our proposed method through a series of experiments. We then apply the framework to analyze neonatal hormone data in Section~4.
Finally, we summarize our findings in Section~5 and discuss some possible directions for future research. 

\section{Methodology: nonlinear MIR}
\subsection{Notation and definition of MI}
Let \( \mathbb{R} \) denote the set of all real numbers and let \( \mathbb{N} \) denote the set of all nonnegative integers.  
For any \( \rho \in \mathbb{N} \cup \{\infty\} \) and open interval \( (a,b) \subseteq \mathbb{R} \) (such that \( a < b \)),  
\( C^{\rho}(a,b) \) denotes the set of all functions that are \( \rho \)-times continuously differentiable on \( (a,b) \)  
and \( \mathrm{cl}(a,b) \) denotes the closure of \( (a,b) \).  
For any \( d \in \mathbb{N} \), let \( \mathbb{P}_d \) be the set of all univariate real polynomials of degree at most \( d \), i.e.,
\(
\mathbb{P}_d \coloneqq \left\{\, 
u : \mathbb{R} \ni x \mapsto \sum_{k=0}^{d} c_k x^{k} \in \mathbb{R} \;\middle|\; c_k \in \mathbb{R}
\,\right\}.
\)
Vectors and matrices are denoted by bold lowercase and uppercase letters, respectively.  

The MI is most effective for strictly unimodal distributions because it represents that interval where the data are most concentrated.
Therefore, in our discussion of the MI, we assume that the underlying distribution is strictly unimodal. 
Under this assumption, the MI for a continuous random variable \(X\) with probability density function \(f_X(x)\) defined on \(x \in (x_{\inf}, x_{\sup})\) is defined as follows \parencite{Yao2023}.

\noindent\textbf{Definition 1 (Modal interval)}:  
For any \( \alpha \in (0,1) \), the MI is defined as

\begin{equation}
\mathrm{MI}_{\alpha} \coloneqq [m_{\alpha}^{\mathrm{L}}, m_{\alpha}^{\mathrm{R}}]
\coloneqq 
\underset{[x_\mathrm{L}, x_\mathrm{R}] \subseteq \mathrm{cl}(x_{\inf}, x_{\sup})}{\arg\min} (x_\mathrm{R} - x_\mathrm{L})
\quad \text{s.t.} \quad
\int_{x_\mathrm{L}}^{x_\mathrm{R}} f_X(x) \, dx = \alpha.
\label{equ:def_mi}
\end{equation}
That is, for a probability density function, we consider intervals the integral of which equals~\(\alpha\).
Among all such intervals, the MI is the shortest because it represents the most concentrated region.

This concept is also known as the shorth \parencite{Einmahl2010}, highest density interval \parencite{Zhu2019}, shortest prediction interval \parencite{Dahiya1982}, shortest modal interval \parencite{DeGooijer2000}, and minimum volume region \parencite{Polonik2000}. 

By extending this concept to the conditional case, the conditional MI \parencite{Yao2023} of \( Y \) given \( X = x \) is defined with respect to the conditional density function \( f_{Y|X}(y \mid x) \) for \( y \in (y_{\inf}, y_{\sup}) \).

\noindent\textbf{Definition 2 (Conditional modal interval)}:  
For any \( \alpha \in (0,1) \), the conditional MI is defined as
\begin{equation}
\begin{aligned}
\mathrm{MI}_{\alpha,Y}(x)
\coloneqq &[m_{\alpha,Y}^{\mathrm{low}}(x), m_{\alpha,Y}^{\mathrm{up}}(x)]\coloneqq
\underset{[y_{\mathrm{low}}, y_{\mathrm{up}}] \subseteq \mathrm{cl}(y_{\inf}, y_{\sup})}{\arg\min} (y_{\mathrm{up}} - y_{\mathrm{low}}) \\
&\text{s.t.} \quad
\int_{y_{\mathrm{low}}}^{y_{\mathrm{up}}} f_{Y|X}(y \mid x) \, dy = \alpha.
\end{aligned}
\label{equ:def_cmi}
\end{equation}

Here, \(m_{\alpha,Y}^{\mathrm{low}}(x)\) and \(m_{\alpha,Y}^{\mathrm{up}}(x)\) denote 
the lower and upper bound functions of the conditional MI, 
while \(y_{\mathrm{low}}\) and \(y_{\mathrm{up}}\) represent their corresponding values at a specific \(x\).
The key objective of the present work is to estimate these two bound functions from the observed data.
Both functions are assumed to be continuous and sufficiently smooth with respect to \(x\)
such that the conditional MI varies smoothly along the domain of \(X\).

\subsection{Limitations of bin-based MIR and motivation}

\begin{table}[tb]
\centering
\caption{Comparison between bin-based MIR and KDE-based MIR}
\label{tab:bin_vs_kde}
\begin{tabular}{lp{4.8cm}p{4.8cm}}
\toprule
\textbf{Property} & \textbf{Bin-based MIR} & \textbf{KDE-based MIR} \\
\midrule
Binning dependence &
Strongly affected by number and width of bins &
Irrelevant to binning \\
\midrule
Conditional MI estimation &
Estimated within bins &
Estimated at each observation using kernel weights \\
\midrule
Continuity &
Quantile levels are constant within bins and independent across bins &
Quantile levels change continuously with \(X\) because kernel weights vary smoothly \\
\midrule
Convergence &
Holds as bin width~\(\to 0\) &
Theoretical convergence under standard kernel assumptions \\
\midrule
Selection strategy &
No principled rule for bin number or width &
Bandwidth chosen in a data-driven way\\
\midrule
Spline setup &
Spline knots tied to bin partitioning &
Spline knots selected independently of MI estimation \\
\bottomrule
\end{tabular}
\end{table}

The idea of MIR proposed by \textcite{Yao2023} was to simplify the complex problem of directly estimating the conditional MI through regression by reformulating it as two simpler subproblems that estimate the upper and lower bounds of the MI separately.
To achieve this, the data were divided into bins along the independent variable \(X\).
Within each bin, the empirical distribution of \(Y\) was obtained and the MI was estimated.
The upper and lower bounds of each MI were then converted into quantile levels, and two polynomial functions were fitted in each bin using a quantile loss function to construct a piecewise regression model, which was smoothly connected across bins to form a spline function.
This approach provided an elegant regression-based solution to the conditional MI estimation.

However, this framework involves several theoretical and practical limitations.
In theory, as the sample size \( n \to \infty \) and the bin width approach zero, the empirical distribution within each bin converges to the conditional distribution \( f_{Y|X}(y \mid x_{\mathrm{mid}}) \) at the bin midpoint \( x_{\mathrm{mid}} \), and the estimated MI converges to the true conditional MI at that point.
In practice, however, the number of bins cannot be arbitrarily large because the total number of coefficients in piecewise polynomial fitting increases with the number of bins, and the bin width therefore cannot shrink to zero.

In addition, each bin becomes an independent local model and the estimated quantile levels of the conditional MI are constant within bins and independent across bins.
Moreover, the conditional distribution of \(Y\) generally changes continuously within each bin, whereas the empirical distribution within the bin treats it as a constant, which leads to biased estimation.
When the sample size is limited, choosing the number of bins becomes difficult given that wide bins ignore within-bin variations, whereas narrow bins yield unstable estimates.
These factors substantially reduce estimation accuracy.

To overcome these problems, the MIR framework must be reformulated.
First, we require a method that does not artificially partition the data into discrete bins when estimating the quantile level parameters required in the quantile loss, so that the quantile levels corresponding to the MI bounds can vary continuously with \(X\).
Second, the selection of spline knots for regression should be independent of how the quantile levels of the conditional MI are estimated.
Finally, we wish to retain the smoothing spline because spline functions allow convenient constraints that prevent the two fitted functions from crossing within the domain to maintain valid intervals even with sparse or complex data.

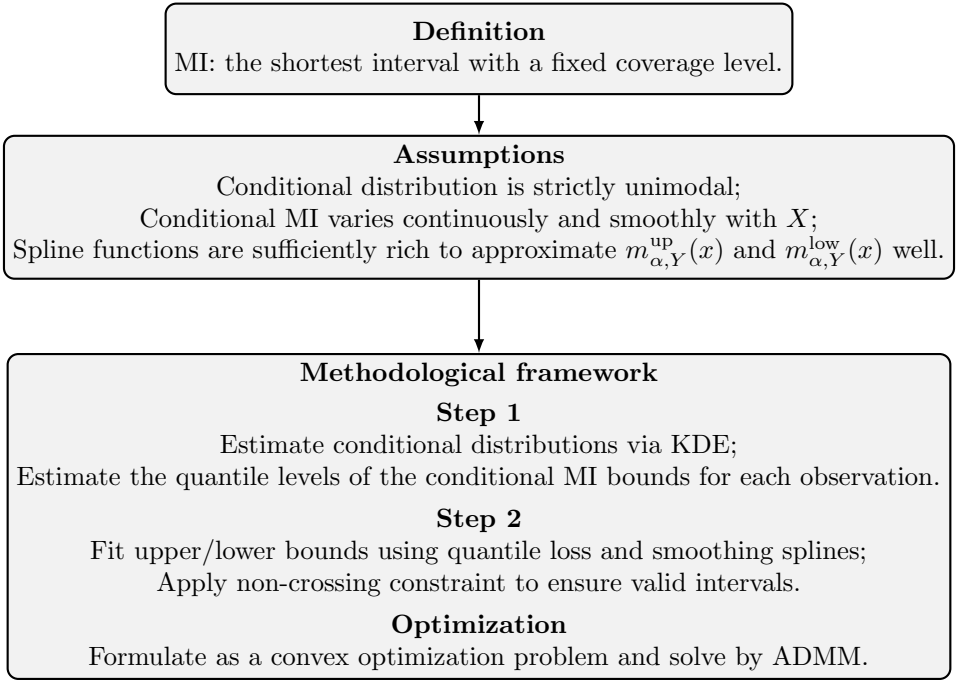
\begin{figure}[t]
\centering
\begin{tikzpicture}[node distance=2.2cm, >=latex, thick]
\tikzstyle{box}=[rectangle, rounded corners, draw=black, align=center,
                 minimum width=6.8cm, minimum height=1.2cm, fill=gray!10]

\node[box] (def) {%
\textbf{Definition}\\
MI: the shortest interval with a fixed coverage level.};

\node[box, below of=def, yshift=0.1cm] (ass) {%
\textbf{Assumptions}\\
Conditional distribution is strictly unimodal;\\
Conditional MI varies continuously and smoothly with \(X\);\\
Spline functions are sufficiently rich to approximate
\(m_{\alpha,Y}^{\mathrm{up}}(x)\) and \(m_{\alpha,Y}^{\mathrm{low}}(x)\) well.};

\node[box, below of=ass, yshift=-1.9cm] (alg) {%
\textbf{Methodological framework}\\[3pt]
\textbf{Step 1} \\
Estimate conditional distributions via KDE;\\
Estimate the quantile levels of the conditional MI bounds for each observation.\\[4pt]
\textbf{Step 2}\\
Fit upper/lower bounds using quantile loss and smoothing splines;\\
Apply non-crossing constraint to ensure valid intervals.\\[4pt]
\textbf{Optimization}\\Formulate as a convex optimization problem and solve by ADMM.};

\draw[->] (def) -- (ass);
\draw[->] (ass) -- (alg);

\end{tikzpicture}
\caption{Logical flow of the proposed MIR framework from definition and assumptions to methodological framework.}
\label{fig:flow}
\end{figure}

Therefore, we reformulate the MIR framework using KDE.
The advantages of KDE are twofold.
First, it eliminates the discontinuities introduced by binning; at any given \(x\), the conditional distribution \( f_{Y|X}(y \mid x) \) and the MI are estimated using smoothly varying kernel weights to ensure that the estimated quantile levels change continuously with \(x\).
Second, bandwidth is determined in a data-driven manner, which avoids the need for arbitrary bin-width selection and does not affect the choice of spline knots.

Accordingly, we propose a two-step reformulated MIR framework.
We refer to it as KDE-based MIR to highlight its difference from bin-based MIR in estimating the quantile level parameters required in the quantile loss.
Figure~\ref{fig:flow} illustrates the logical flow of the proposed framework based on the theoretical definition and assumptions required by the implementation.
In Step 1, KDE is used to estimate, at each observation, the quantile levels corresponding to the conditional MI bounds.
In Step~2, the upper and lower bounds are estimated simultaneously using a quantile loss function with smoothing splines.

Table~\ref{tab:bin_vs_kde} summarizes the main conceptual and theoretical differences between the bin- and KDE-based MIR frameworks. The comparison highlights how the proposed reformulation achieves continuity in estimation, data-driven adaptability, and theoretical convergence.

\subsection{Formulation of the nonlinear MIR}
\subsubsection{Step 1: pointwise MI quantile level estimation}
Let the observed data comprise a set of \(n\) pairs \(\{(x_i, y_i)\}_{i=1}^{n}\), where \(x_i\) and \(y_i\) denote the \(i\)th observations of the independent and dependent variables, respectively.
By definition, the estimation of the conditional MI depends on the conditional cumulative distribution of \(Y\) given \(X=x\).
Therefore, the first step is to estimate this conditional cumulative distribution for each \(x\).

KDE is one of the most commonly used methods to estimate the conditional MI \parencite{DeGooijer2000,Polonik2000}.
In our method, the Nadaraya–Watson estimator \parencite{Li2007} is employed because it is computationally efficient, which is important when the MI needs to be evaluated at many observed points.
The conditional cumulative distribution of \(Y\) given \(X=x\) is estimated as
\begin{equation}
\hat{F}_{Y|x}(y)
= \frac{\sum_{i=1}^{n} K\!\left(\frac{x_i - x}{h}\right) \mathbf{1}(y_i \le y)}
{\sum_{i=1}^{n} K\!\left(\frac{x_i - x}{h}\right)},
\label{equ:kde_cdf}
\end{equation}
where \( K(\cdot) \) is the kernel function (the Gaussian kernel is used in this study),  and
\( h \) is the bandwidth of \(X\), which controls the smoothness of the kernel density estimator.  
The bandwidth was selected using a plug-in method \parencite{Sheather1991,Botev2010,Li2007}.  
\( \mathbf{1}(y_i \le y) \) is an indicator function that equals 1 if \( y_i \le y \) and 0 otherwise.

After determining the coverage level \(\alpha \in (0,1)\), 
we estimate the \(100\alpha\%\) conditional MI defined in Equation~(\ref{equ:def_cmi}) 
for each observation \((x_i, y_i)\) based on the estimated conditional distribution \(\hat{F}_{Y|x}(y)\) as follows.

\begin{equation}
\label{equ:kde_cmi}
\begin{aligned}
\widehat{\mathrm{MI}}_{\alpha,Y}(x_i)
\coloneqq &[\hat{m}_{\alpha,Y}^{\mathrm{low}}(x_i), \hat{m}_{\alpha,Y}^{\mathrm{up}}(x_i)] \coloneqq
\underset{[\hat{y}_{\mathrm{low}}, \hat{y}_{\mathrm{up}}] \subseteq \mathrm{cl}(y_{\inf}, y_{\sup})}{\arg\min}
(\hat{y}_i^{\mathrm{up}} - \hat{y}_i^{\mathrm{low}}) \\
&\text{s.t.} \quad
\hat{F}_{Y|x_i}(\hat{y}_i^{\mathrm{up}}) - \hat{F}_{Y|x_i}(\hat{y}_i^{\mathrm{low}}) \ge \alpha.
\end{aligned}
\end{equation}

Assuming that \( f_{Y|X}(y \mid x) \) is strictly unimodal  
the conditional quantile function of \( Y \) given \( X = x \) is defined as
\[
Q_{Y|x}(p) \coloneqq F_{Y|x}^{-1}(p), \quad \text{for } p \in (0,1),
\]
where \(p\) is the quantile level.
The core idea of our method is to use Equation~\eqref{equ:kde_cmi} to estimate, at each observed covariate value \(x_i\), the quantile levels correspoding to the conditional MI bounds, denoted by
\(
\ubar{p}_i \coloneqq \hat{F}_{Y\mid x_i}(\hat{y}_i^{\mathrm{low}}),\)
and \(\bar{p}_i \coloneqq \hat{F}_{Y\mid x_i}(\hat{y}_i^{\mathrm{up}}).\)
These quantile levels serve as the quantile level parameters in the quantile loss in Step~2, where smoothing splines are fitted to simultaneously regress the lower and upper bound curves of the conditional MI.

\subsubsection{Step 2: simultaneous regression of MI bounds}
We employ the quantile loss function defined as
\[
\mathcal{J}_p(t) \coloneqq
\begin{cases}
    p t, & \text{if } t \ge 0, \\
    -(1 - p)t, & \text{if } t < 0,
\end{cases}
\]
which has the property that its expected value is minimized at the conditional \(p\)th quantile of \( Y \) given \( X = x \) \parencite{Koenker1978,Koenker2005}.
\begin{equation}
F^{-1}_{Y|x}(p)
= \arg\min_{q} \, \mathbb{E}\!\left[ \mathcal{J}_p(Y - q) \mid X = x \right].
\end{equation}
The sample quantile minimizes the empirical quantile loss and converges in probability to the true quantile under suitable regularity conditions \parencite{Koenker1978,Koenker2005}.

Based on this property, if the quantile level \(p\) is set to a value corresponding to the upper or lower bound of the conditional MI, each bound can be theoretically expressed as the minimizer of the expected quantile loss.
\begin{equation}
m_{\alpha,Y}^{\mathrm{up}}(x) =
\arg\min_{q} \, \mathbb{E} \left[
\mathcal{J}_{F_{Y \mid x}(m_{\alpha,Y}^{\mathrm{up}}(x))}(Y - q) \mid X = x
\right],
\label{equ:asym_loss_upper_2d}
\end{equation}

\begin{equation}
m_{\alpha,Y}^{\mathrm{low}}(x) =
\arg\min_{q} \, \mathbb{E} \left[
\mathcal{J}_{F_{Y \mid x}(m_{\alpha,Y}^{\mathrm{low}}(x))}(Y - q) \mid X = x
\right].
\label{equ:asym_loss_lower_2d}
\end{equation}

In practice, however, the true quantile levels corresponding to the upper and lower bounds are unknown because the conditional distribution \(F_{Y \mid x}\) cannot be observed directly.
Therefore, in Step~1, we estimate the quantile levels corresponding to the upper and lower bounds for each \(x_i\) using KDE, which are denoted by \(\overline{p}_{i}\) and \(\ubar{p}_{i}\), respectively.

We use spline functions because they not only provide high flexibility and smoothness but also allow easy control over the curve shape \parencite{Schumaker2015}.
Spline knots are denoted as \(\xi_j\), with \(\xi_0 < \xi_1 < \cdots < \xi_b\).
The intervals between knots are defined as \(I_j \coloneqq [\xi_{j-1}, \xi_j]\) \((j = 1, 2, \cdots, b)\).
For a partition \(\sqcup_b \coloneqq \{I_j\}_{j=1}^b\) and integers \(\rho, d \in \mathbb{N}\) satisfying \(0 \le \rho \le d\), we define
\[
\mathcal{S}_d^\rho(\sqcup_b) \coloneqq 
\{ s \in C^\rho(\xi_0, \xi_b) \mid s = u_j \in \mathbb{P}_d \text{ on } I_j \in \sqcup_b \},
\]
where \(\mathcal{S}_d^\rho(\sqcup_b)\) denotes the set of univariate spline functions of degree at most \(d\) and smoothness \(\rho\) defined on the partition \(\sqcup_b\).
The domain \(I\) covers the range of the data, i.e., \(I \supseteq (x_{\min}, x_{\max}) \coloneqq (\min\{x_i\}, \max\{x_i\})\).

The upper bound is estimated using a smoothing spline with the quantile loss function as follows.
\begin{equation}
\underset{\bar{s} \in \mathcal{S}_d^\rho(\sqcup_b)}{\text{minimize}}
\sum_{i=1}^n w_i \mathcal{J}_{\bar{p}_i}(y_i - \bar{s}(x_i))
+ \lambda \int_I |\bar{s}''(x)|^2 \, \mathrm{d}x ,
\label{equ:upper_opt}
\end{equation}
where \(\bar{s}(x)\) is a spline curve approximating \( m_{\alpha,Y}^{\mathrm{up}}(x) \),
\(w_i > 0\) are the observation weights, and \(\lambda > 0\) is the smoothing parameter.
Each observation has its own quantile level \(\bar{p}_i\), unlike in standard quantile regression.

By standard convergence properties of kernel estimators under regularity conditions, as \( n \to \infty\), \( h \to 0\), and \( nh \to \infty\), the estimated conditional distribution in Equation~\eqref{equ:kde_cdf} converges almost surely to the true conditional distribution \parencite{Li2007}.
Consequently, the quantile levels \(\hat{F}_{Y|x}(\hat{y}_{\mathrm{up}})\) and \(\hat{F}_{Y|x}(\hat{y}_{\mathrm{low}})\) estimated via Equation~\eqref{equ:kde_cmi} converge almost surely to their true quantile levels
\(F_{Y|x}(y_{\mathrm{up}})\) and \(F_{Y|x}(y_{\mathrm{low}})\), respectively.

By properties of the quantile loss and Equation~\eqref{equ:asym_loss_upper_2d}, when the quantile level $p$ is set to the true quantile level corresponding to the upper bound, the minimizer of the empirical quantile loss converges almost surely to the true upper bound as \(n\to\infty\) under regularity conditions \parencite{Koenker1978,Koenker2005}.

We further assume that the smoothing spline representation is sufficiently expressive so that \(\bar{s}(x)\) can exactly represent \(m_{\alpha,Y}^{\mathrm{up}}(x)\) for an appropriate choice of coefficients.
Combining the convergence properties of KDE and the quantile loss, it follows that as \(n\to\infty\), \(h\to 0\), and \(nh\to\infty\), the solution \(\bar{s}^*(x)\) of Equation~\eqref{equ:upper_opt} converges almost surely to \(m_{\alpha,Y}^{\mathrm{up}}(x)\).

Similarly, using the quantile levels \(\ubar{p}_i\), 
the lower bound \( m_{\alpha,Y}^{\mathrm{low}}(x) \) can be estimated by \(\ubar{s}(x)\).

However, estimating the upper and lower bounds separately may result in \(\bar{s}^*(x) < \ubar{s}^*(x)\) at some \(x\), especially when the data are limited or complex.
To prevent this, we combine the optimization problems for the upper and lower bounds and introduce a non-crossing constraint.

\begin{equation}
\begin{aligned}
\underset{\bar{s},\,\ubar{s} \in \mathcal{S}_d^\rho(\sqcup_b)}{\text{minimize}} \ 
 \sum_{i=1}^n w_i \mathcal{J}_{\bar{p}_i}(y_i - \bar{s}(x_i))&
+ \lambda \int_I |\bar{s}''(x)|^2 \, \mathrm{d}x +\sum_{i=1}^n w_i \mathcal{J}_{\ubar{p}_i}(y_i - \ubar{s}(x_i))
+ \lambda \int_I |\ubar{s}''(x)|^2 \, \mathrm{d}x \\
& \text{subject to}\quad \forall x \in I\quad \bar{s}(x) \ge \ubar{s}(x).
\end{aligned}
\label{equ:joint_opt}
\end{equation}

In practice, the observation weights \(w_i\) are determined by the data density to improve estimation stability \parencite{Yao2023}.
We set \(w_i = \hat{f}(x_i)^{1/5}\), where \(\hat{f}(x_i)\) is the KDE-based density estimate \parencite{Li2007}.
\begin{equation}
\label{equ:kde_x}
\hat{f}(x_i) \coloneqq \frac{1}{nh} \sum_{l=1}^n K\!\left(\frac{x_i - x_l}{h}\right).
\end{equation}
This density weighting assigns higher weights to regions with more data to ensure stable estimates, whereas low-density areas receive smaller weights to mitigate overfitting.

Note that \(\hat{f}(x_i)\) can vary substantially across the domain, and using \(\hat{f}(x_i)\) itself may lead to overly small weights in low-density areas.
We therefore use the power transform 
\(\hat{f}(x_i)^{1/5}\) to compress the weight range while preserving the ordering, so that low-density areas are downweighted but not ignored.
A larger exponent puts more emphasis on high-density areas, whereas a smaller exponent yields weights closer to uniform.

The smoothing parameter \(\lambda\) controls the balance between fidelity and smoothness to prevent overfitting.
Applying a common smoothing parameter \(\lambda\) for both because the upper and lower bounds are estimated from the same dataset and share similar smoothness properties.
This choice ensures consistent smoothness between the two bounds and simplifies model selection by avoiding separate tuning.
It also follows previous studies that adopted a single smoothing parameter \parencite[e.g.][]{Silverman1985,Kitahara2020,Yao2023}.

In practice, cubic splines (\(d = 3, \rho = 2\)) with uniformly spaced knots generally suffice.
When the number of knots exceeds a certain threshold, the specific placement of knots becomes less important, as the smoothness is primarily governed by \(\lambda\) \parencite{Ruppert2002}.
Therefore, in this study, we uniformly select a sufficiently large number of knots and adjust the smoothness through \(\lambda\).

\subsubsection{Smoothing parameter selection}
To select an appropriate smoothing parameter \( \lambda \), 
we improve the coverage width-based criterion (CWC), which was originally proposed to estimate prediction intervals using NNs \parencite{Khosravi2011}.
We propose a modified version called the mCWC, which ensures the continuity of the penalty term at \(\mathrm{MICP}=\alpha\) 
and is thus more suitable for the MIR framework. 
Here, MICP denotes the modal interval coverage probability, i.e., the proportion of test data points the true values of which fall within the estimated MI. 
Specifically, the mCWC combines the normalized mean modal interval width (NMMIW) with MICP to identify the shortest interval satisfying the target coverage level \(\alpha\) as given below.

\begin{equation}
\label{eq:mCWC}
\mathrm{mCWC} \coloneqq
\begin{cases}
\mathrm{NMMIW}, & \text{if } \mathrm{MICP} \geq \alpha, \\
\mathrm{NMMIW} \cdot \exp[-\eta(\mathrm{MICP} - \alpha)], & \text{otherwise}.
\end{cases}
\end{equation}

Here, \( \mathrm{NMMIW} \) represents the mean width of the conditional MIs at the test points normalized by the range of the response variable \(Y\).
The mCWC criterion can be viewed as a tradeoff between interval width and coverage: we aim to obtain short intervals while meeting the target coverage level \(\alpha\).
When \( \mathrm{MICP}<\alpha \), we apply an exponential penalty to \( \mathrm{NMMIW} \) to avoid selecting models whose coverage is far below the target level \( \alpha \).

\begin{figure}[t]
    \centering
    \includegraphics[width=0.5\linewidth]{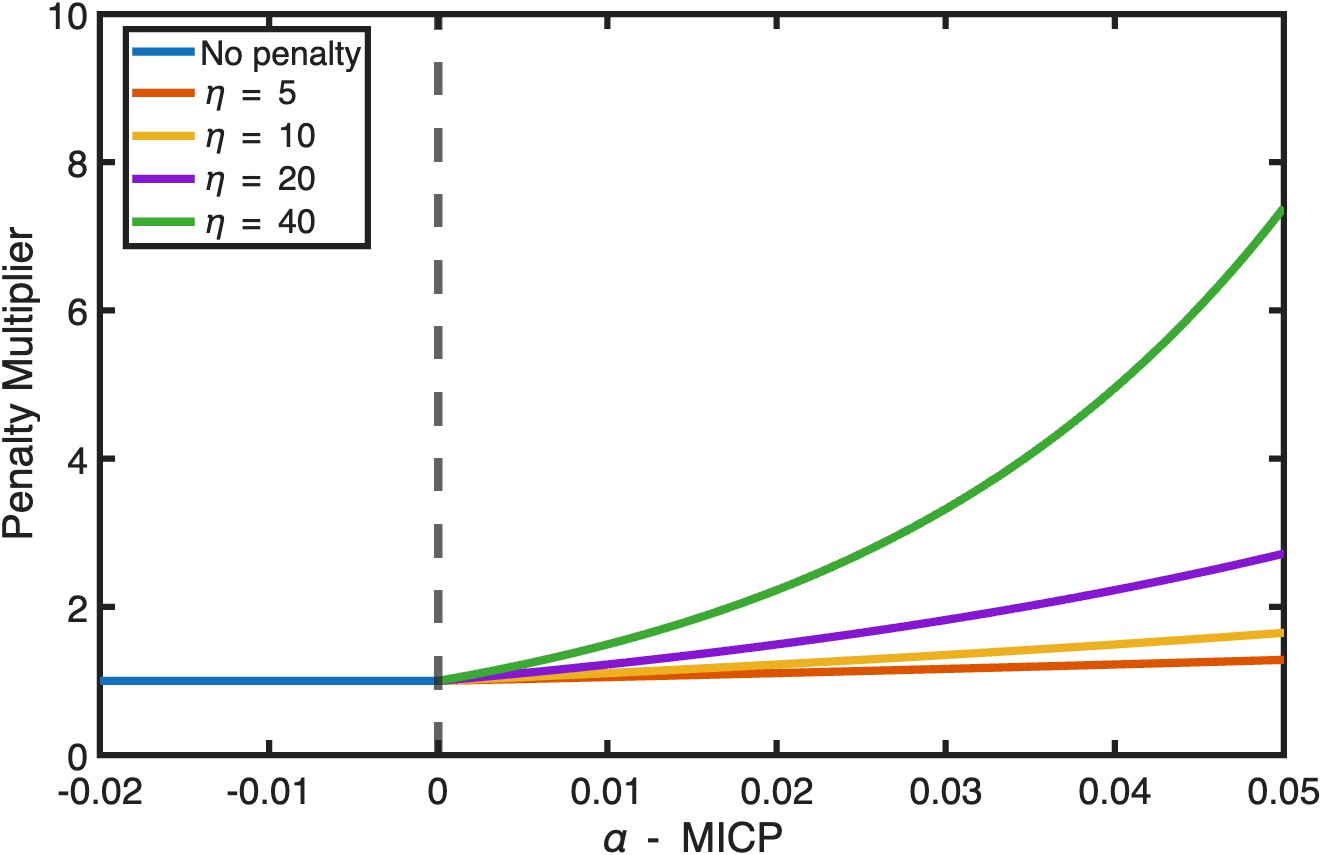}
    \caption{Penalty multiplier in the mCWC criterion as a function of \(\alpha - \mathrm{MICP}\).
When \(\mathrm{MICP}\geq \alpha\), the multiplier equals 1 (no penalty).
When 
\(\mathrm{MICP}< \alpha\), the multiplier increases as \(\exp[-\eta(\mathrm{MICP}-\alpha)]\) for \(\eta\in\{5,10,20,40\}\), yielding larger penalties as coverage falls further below the target. The dashed vertical line marks \(\mathrm{MICP}= \alpha\).}
    \label{fig:mcwc_eta}
\end{figure}

The parameter \( \eta \) controls the strength of this penalty and thus determines the tradeoff between interval width and under-coverage: larger \( \eta \) indicates less tolerance for \( \mathrm{MICP}<\alpha \).
Figure~\ref{fig:mcwc_eta} illustrates how the penalty multiplier increases with different values for \( \eta\in\{5,10,20,40\} \).
In this study, we set \(\eta=20\) to discourage models with clearly insufficient coverage, while still allowing models with slightly lower-than-$\alpha$ coverage to be selected when they yield substantially shorter intervals.

In practice, mCWC is used as a model-selection criterion to choose \(\lambda\) from a prespecified candidate set.
Specifically, we first specify a candidate set \(\Lambda=\{\lambda_1,\ldots,\lambda_{\ell}\}\) and compute \(\mathrm{mCWC}(\lambda)\) using \(V\)-fold cross-validation.
For each \(\lambda\in\Lambda\), we fit the proposed MIR model on the training folds to obtain the estimated lower and upper bound functions of the conditional MI.
The region between these two fitted curves is taken as the estimated conditional MI.
We then compute \(\mathrm{NMMIW}(\lambda)\) and \(\mathrm{MICP}(\lambda)\) on the held-out fold, and hence \(\mathrm{mCWC}(\lambda)\) via Equation~\eqref{eq:mCWC}.
Repeating this over all folds yields \(V\) mCWC values for each \(\lambda\), and we use their average, denoted by \(\overline{\mathrm{mCWC}}(\lambda)\), as the selection score.
Finally, we select the smoothing parameter by
\[
\lambda^{\ast} \coloneqq \arg\min_{\lambda\in\Lambda} \overline{\mathrm{mCWC}}(\lambda).
\]

\subsection{Vector formulation}
We represent each spline function \(s(x)\) in a normalized interval form as follows.
\begin{equation}
s(x) = u_j(x) = c_0^{\langle j \rangle} + \sum_{k=1}^{d} c_k^{\langle j \rangle} \left( \frac{x - \xi_{j-1}}{\Delta_j} \right)^k, 
\quad \text{for } x \in I_j,
\end{equation}
where \(c_k^{\langle j \rangle} \in \mathbb{R}\) \((k = 0, \ldots, d)\) represents the \(k\)th coefficient of the \(j\)th polynomial segment \(u_j \in \mathbb{P}_d\) \((j = 1, \ldots, b)\), and \(\Delta_j \coloneqq \xi_j - \xi_{j-1}\) denotes the length of the interval \(I_j\).
The coefficient vector of \(s(x)\) is then defined as
\[
\tilde{\bm{c}} \coloneqq (c_0^{\langle 1 \rangle}, c_1^{\langle 1 \rangle}, \ldots, c_d^{\langle 1 \rangle}, 
c_0^{\langle 2 \rangle}, c_1^{\langle 2 \rangle}, \ldots, c_d^{\langle 2 \rangle}, \ldots, c_d^{\langle b \rangle})^{\mathrm{T}} 
\in \mathbb{R}^{b(d+1)}.
\]
Given that \(s \in C^{\rho}(\xi_0, \xi_b)\), the coefficients of adjacent polynomial segments \(u_j \in \mathbb{P}_d\) and \(u_{j+1} \in \mathbb{P}_d\) \((j = 1, \ldots, b-1)\) need to meet the following linear conditions.
\begin{equation}
\frac{1}{\Delta_j^g} \sum_{k=g}^{d} \frac{k!}{(k-g)!} c_k^{\langle j \rangle}
- \frac{g!}{\Delta_{j+1}^g} c_g^{\langle j+1 \rangle} = 0
\quad \Leftrightarrow \quad
u_j^{(g)}(\xi_j) = u_{j+1}^{(g)}(\xi_j),
\quad g = 0, 1, \ldots, \rho,
\label{equ:continous}
\end{equation}
where \(u_j^{(g)}(x)\) denotes the \(g\)th derivative of \(u_j(x)\). From Equation~(\ref{equ:continous}), we can define a matrix as 
\(\tilde{\bm{H}} \in \mathbb{R}^{(b-1)(\rho+1) \times b(d+1)}\) such that 
\(\tilde{\bm{H}} \tilde{\bm{c}} = \mathbf{0} \Leftrightarrow s \in \mathcal{S}_d^{\rho}(\sqcup_b)\),
where \(\mathbf{0}\) is the zero vector.

First, we represent the loss function in Equation~(\ref{equ:joint_opt}) using \(\tilde{\bm{c}}\). For each observation \(x_i \in I_j\), the function value is given by \(s(x_i) = \bm{a}_i^{\mathrm{T}} \tilde{\bm{c}}\), where the vector \(\bm{a}_i \in \mathbb{R}^{b(d+1)}\) is defined as
\[
\bm{a}_i \coloneqq 
(\underbrace{0, 0, \ldots, 0}_{(j-1)(d+1)}, 
1, \tfrac{x_i - \xi_{j-1}}{\Delta_j}, \ldots, 
(\tfrac{x_i - \xi_{j-1}}{\Delta_j})^d, 
\underbrace{0, 0, \ldots, 0}_{(b-j)(d+1)})^{\mathrm{T}}.
\]
Therefore, we obtain \(\mathcal{J}_p(y_i - s(x_i)) = \mathcal{J}_p(y_i - \bm{a}_i^{\mathrm{T}} \tilde{\bm{c}})\).

Next, we represent the roughness penalty in Equation~(\ref{equ:joint_opt}) using \(\tilde{\bm{c}}\). The total roughness penalty on \(I\) can be decomposed into the sum of penalties over the subintervals \(I_j\).
\begin{equation}
\int_I |s''(x)|^2 \, dx = \sum_{j=1}^{b} \int_{I_j} |u_j''(x)|^2 \, dx.
\label{equ:pen_pol}
\end{equation}
For each subinterval \(I_j\), this penalty can be expressed as a quadratic form:
\begin{equation}
\int_{I_j} |u_j''(x)|^2 \, dx 
= \sum_{k=2}^{d} \sum_{g=2}^{d} 
\frac{k(k-1)g(g-1)}{\Delta_j^{3}(k+g-3)} 
\, c_k^{\langle j \rangle} c_g^{\langle j \rangle}.
\label{equ:pen_pol_detail}
\end{equation}
From Equations~(\ref{equ:pen_pol})~ and~(\ref{equ:pen_pol_detail}), there exists a symmetric positive semidefinite matrix 
\(\tilde{\bm{Q}} \in \mathbb{R}^{b(d+1) \times b(d+1)}\) such that
\[
\int_I |s''(x)|^2 \, dx = \tilde{\bm{c}}^{\mathrm{T}} \tilde{\bm{Q}} \tilde{\bm{c}}.
\]

Finally, we represent the linear non-crossing constraint in Equation~(\ref{equ:joint_opt}) using the coefficient vector.
Although deriving a practically necessary and sufficient condition for non-crossing is complex, \textcite{Yao2023} extended the sufficient condition for nonnegativity proposed by \textcite{Hess1994} such that it can be applied to ensure the non-crossing of the spline function \(s(x)\) over \(I_j\).
The nonnegativity constraint is given as follows.
\begin{equation}
\sum_{k=0}^{g} \frac{(d-k)!}{(d-g)!(g-k)!} c_k^{\langle j \rangle} \ge 0, 
\quad g = 0, 1, \ldots, d
\;\Rightarrow\;
s(x) \ge 0 \text{ for all } x \in I_j.
\label{equ:nonnegativity}
\end{equation}

We define \(\bar{\bm{c}} \coloneqq (\bar{c}_0^{\langle 1 \rangle}, \bar{c}_1^{\langle 1 \rangle}, \ldots, \bar{c}_d^{\langle b \rangle})^{\mathrm{T}} \in \mathbb{R}^{b(d+1)}\) and 
\(\ubar{\bm{c}} \coloneqq (\ubar{c}_0^{\langle 1 \rangle}, \ubar{c}_1^{\langle 1 \rangle}, \ldots, \ubar{c}_d^{\langle b \rangle})^{\mathrm{T}} \in \mathbb{R}^{b(d+1)}\) 
as the coefficient vectors for \(\bar{s}(x)\) and \(\ubar{s}(x)\), respectively.  
The non-crossing constraint \(\forall x \in I, \; \bar{s}(x) - \ubar{s}(x) \ge 0\) is equivalent to ensuring the nonnegativity of the spline function \(\bar{s}(x) - \ubar{s}(x)\).  
The coefficients of the \(j\)th polynomial segment of \(\bar{s}(x) - \ubar{s}(x)\) are given by 
\((\bar{c}_0^{\langle j \rangle} - \ubar{c}_0^{\langle j \rangle}, \bar{c}_1^{\langle j \rangle} - \ubar{c}_1^{\langle j \rangle}, \ldots, \bar{c}_d^{\langle j \rangle} - \ubar{c}_d^{\langle j \rangle})^{\mathrm{T}}\).  
Thus, we obtain the sufficient condition from Equation~(\ref{equ:nonnegativity}).
\begin{equation}
\sum_{k=0}^{g} \frac{(d-k)!}{(d-g)!(g-k)!} 
(\bar{c}_k^{\langle j \rangle} - \ubar{c}_k^{\langle j \rangle}) \ge 0,
\quad g = 0, 1, \ldots, d
\;\Rightarrow\;
\bar{s}(x) \ge \ubar{s}(x) \text{ for all } x \in I_j.
\label{equ:noncrossing}
\end{equation}
From Equation~(\ref{equ:noncrossing}), there exists a matrix 
\(\tilde{\bm{G}} \in \mathbb{R}^{b(d+1) \times b(d+1)}\) such that 
\(\tilde{\bm{G}}(\bar{\bm{c}} - \ubar{\bm{c}}) \ge \mathbf{0}
\Rightarrow \bar{s}(x) \ge \ubar{s}(x)\) for all \(x \in I\).

We define 
\(
\bm{w} \coloneqq (w_1, \ldots, w_n, w_1, \ldots, w_n)^{\mathrm{T}} \in \mathbb{R}^{2n}, \quad
\bm{p} \coloneqq (\bar{p}_1, \ldots, \bar{p}_n, \ubar{p}_1, \ldots, \ubar{p}_n)^{\mathrm{T}} \in \mathbb{R}^{2n}
\),
\(
\bm{y} \coloneqq (y_1, \ldots, y_n, y_1, \ldots, y_n)^{\mathrm{T}} \in \mathbb{R}^{2n}
\) and
\(
\bm{c} \coloneqq (\bar{\bm{c}}^{\mathrm{T}}, \ubar{\bm{c}}^{\mathrm{T}})^{\mathrm{T}} \in \mathbb{R}^{2b(d+1)}
\).
Let 
\(\tilde{\bm{A}} \coloneqq (\bm{a}_1, \ldots, \bm{a}_n)^{\mathrm{T}} \in \mathbb{R}^{n \times b(d+1)}\),
and consider the block matrix
\(
\bm{A} \coloneqq 
\begin{pmatrix}
\tilde{\bm{A}} & \bm{O} \\
\bm{O} & \tilde{\bm{A}}
\end{pmatrix}
\in \mathbb{R}^{2n \times 2b(d+1)},
\)
where \(\bm{O}\) is the zero matrix.
The two loss terms in Equation~(\ref{equ:joint_opt}) can be jointly written as a convex function \(\mathcal{J}_{\bm{p}, \bm{y}, \bm{w}}(\bm{A}\bm{c}) : \mathbb{R}^{2n} \to [0, \infty)\) that admits a proximal operator defined as
\[
\mathcal{J}_{\bm{p}, \bm{y}, \bm{w}}(\bm{A}\bm{c}) 
\coloneqq 
\bm{w}^{\mathrm{T}}
\bigl(
\mathcal{J}_{\bar{p}_1}(y_1 - \bm{a}_1^{\mathrm{T}} \bar{\bm{c}}),
\ldots,
\mathcal{J}_{\bar{p}_n}(y_n - \bm{a}_n^{\mathrm{T}} \bar{\bm{c}}),
\ldots,
\mathcal{J}_{\ubar{p}_n}(y_n - \bm{a}_n^{\mathrm{T}} \ubar{\bm{c}})
\bigr)^{\mathrm{T}}.
\]

Similarly, we define the block matrices
\(
\bm{H} \coloneqq 
\begin{pmatrix}
\tilde{\bm{H}} & \bm{O} \\
\bm{O} & \tilde{\bm{H}}
\end{pmatrix}
\in \mathbb{R}^{2(b-1)(\rho+1) \times 2b(d+1)}
\),
\(
\bm{Q} \coloneqq 
\begin{pmatrix}
\tilde{\bm{Q}} & \bm{O} \\
\bm{O} & \tilde{\bm{Q}}
\end{pmatrix}
\in \mathbb{R}^{2b(d+1) \times 2b(d+1)},
\) and
\(
\bm{G} \coloneqq 
\begin{pmatrix}
\tilde{\bm{G}} & -\tilde{\bm{G}}
\end{pmatrix}
\in \mathbb{R}^{b(d+1) \times 2b(d+1)}
\).
This allows us to express the \(\rho\)-times continuous differentiability of \(\bar{s}(x)\) and \(\ubar{s}(x)\) as \(\bm{H}\bm{c} = \mathbf{0}\),  
the two roughness penalty terms as \(\lambda \bm{c}^{\mathrm{T}} \bm{Q}\bm{c}\),  
and the non-crossing constraint as \(\bm{G}\bm{c} \ge \mathbf{0}\).

Finally, the optimization problem in Equation~(\ref{equ:joint_opt}) is reformulated into the following convex optimization form.
\begin{equation}
\begin{aligned}
&\underset{\bm{c}}{\text{minimize}} \quad 
 \mathcal{J}_{\bm{p}, \bm{y}, \bm{w}}(\bm{A}\bm{c}) + \lambda \bm{c}^{\mathrm{T}} \bm{Q}\bm{c}, \\
&\text{subject to} \quad 
 \bm{H}\bm{c} = \mathbf{0}, \quad \bm{G}\bm{c} \ge \mathbf{0}.
\end{aligned}
\label{equ:convex}
\end{equation}

\subsection{Convex optimization}
We employed the alternating direction method of multipliers (ADMM) to solve the convex optimization problem in Equation~(\ref{equ:convex}).
ADMM is particularly well suited to this problem because it decomposes complex optimization tasks into smaller, more manageable subproblems \parencite{Boyd2011,Glowinski2014,Condat2023}.
ADMM handles both constraints and large-scale structures efficiently by introducing auxiliary variables.

We introduce two auxiliary vectors \(\bm{z}_1 \in \mathbb{R}^{2n}\) and \(\bm{z}_2 \in \mathbb{R}^{b(d+1)}\). This leads to the following ADMM formulation.
\begin{equation}
\begin{aligned}
&\underset{\bm{c}, \bm{z}_1, \bm{z}_2}{\text{minimize}}\quad
 \lambda \bm{c}^{\mathrm{T}} \bm{Q}\bm{c} 
+ \iota_{=0}(\bm{H}\bm{c}) 
+ \mathcal{J}_{\bm{p}, \bm{y}, \bm{w}}(\bm{z}_1)
+ \iota_{\ge 0}(\bm{z}_2), \\
&\text{subject to} \quad 
 \bm{z}_1 = \bm{A}\bm{c}, \quad 
\bm{z}_2 = \bm{G}\bm{c},
\end{aligned}
\end{equation}
where the indicator function \(\iota_{=0}(\bm{H}\bm{c})\) equals 0 when the constraint \(\bm{H}\bm{c} = \mathbf{0}\) is satisfied and takes the value \(+\infty\) otherwise. Similarly, \(\iota_{\ge 0}(\bm{z}_2)\) equals 0 if \(\bm{z}_2 \ge \mathbf{0}\), and \(+\infty\) otherwise.

For every \(\bm{z} = (z_1, z_2, \ldots, z_{2n})^{\mathrm{T}} \in \mathbb{R}^{2n}\),  
the proximity operator \(\mathrm{prox}_{\mathcal{J}_{\bm{p}, \bm{y}, \bm{w}}} : \mathbb{R}^{2n} \to \mathbb{R}^{2n}\)  
of the proximable convex function \(\mathcal{J}_{\bm{p}, \bm{y}, \bm{w}}(\bm{z})\) is computed component-wise as
\[
\mathrm{prox}_{w \mathcal{J}_p (y - \cdot)}(z) =
\begin{cases}
z + \gamma p w, & \text{if } y - z \ge \gamma p w, \\[0.4em]
z - \gamma (1 - p) w, & \text{if } y - z \le -\gamma (1 - p) w, \\[0.4em]
y, & \text{otherwise},
\end{cases}
\]
where \(\gamma > 0\) is the ADMM parameter; in the proposed method, we set \(\gamma = 1\) for simplicity.
Similarly, for every \(\bm{z} = (z_1, z_2, \ldots, z_{b(d+1)})^{\mathrm{T}} \in \mathbb{R}^{b(d+1)}\),  
the proximity operator \(\mathrm{prox}_{\iota_{\ge 0}} : \mathbb{R}^{b(d+1)} \to \mathbb{R}^{b(d+1)}\)  
of \(\iota_{\ge 0}(\bm{z})\) is given by the projection operator
\[
\mathrm{prox}_{\iota_{\ge 0}}(z) =
\mathrm{projection}_{\ge 0}(z) =
\begin{cases}
z, & \text{if } z \ge 0, \\
0, & \text{if } z < 0.
\end{cases}
\]

\section{Numerical experiments}
In this section, we describe numerical experiments conducted to evaluate the performance of the proposed KDE-based MIR method.
In contrast to a confidence interval, which aims to capture true parameter values with a high coverage probability (e.g., 90\% or 95\%), 
the MI represents the most concentrated interval of the data.
We set the coverage level to 0.5, referring to the proportion of data between the first and third quartiles, which also includes half of the data.

The 50\% MI was then estimated using three methods, including the KDE method (Equation~(\ref{equ:kde_cmi})), the bin-based MIR method \parencite{Yao2023}, and the proposed KDE-based MIR method (Equation~(\ref{equ:joint_opt})). All simulations and computations were performed using MATLAB 2023b.
The MATLAB implementation of the proposed KDE-based MIR method is publicly available on GitHub \parencite{yao_mir_matlab_github}.

We generated 50 training datasets from each of the two simulated distributions, with each dataset containing a prespecified number of observations.
Specifically, we consider training sample sizes \(n\in\{500,1000,2000,3000\}\).
Since the data-generating mechanisms are fully known in the simulation setting, the true conditional MI is available.
Therefore, we do not select the smoothing parameter \(\lambda\) via the mCWC criterion; instead, we evaluate all candidate \(\lambda\) values and report their performance.

Estimation accuracy was evaluated using three metrics: the root mean squared error (RMSE), the coverage probability (CP), and the average interval width (AIW).
The RMSE quantifies the discrepancy between the estimated and true MI bounds.
To compute the RMSE, we uniformly selected 101 covariate points and compared the estimated intervals with the corresponding true intervals at these locations.
In addition, an independently generated test dataset of size 1{,}000 was used to evaluate predictive performance in terms of CP and AIW.
Note that the mCWC can be computed from CP and AIW.
In the simulation setting, we further use the RMSE to identify the best-performing \(\lambda\) when needed.

Section~\ref{sec:3-3} investigates the effect of the training sample size on estimation accuracy and stability.
By repeating the experiments for \(n\in\{500,1000,2000,3000\}\), we assess how performance improves as more data become available and evaluate robustness under smaller-sample regimes.

Section~\ref{sec:3-4} examines the effect of the smoothing parameter \(\lambda\).
By reporting the results across all candidate \(\lambda\) values, we illustrate how \(\lambda\) affects estimation performance.

Section~\ref{sec:3-5} reports the computation time for each training sample size under the same implementation environment. This provides an empirical characterization of how the computational cost of the proposed method grows with \(n\), and helps assess feasibility for larger datasets.

\subsection{Simulated distributions}
\begin{figure}[tbp]
    \centering
    \begin{subfigure}{0.48\textwidth}
        \centering
        \includegraphics[width=\linewidth]{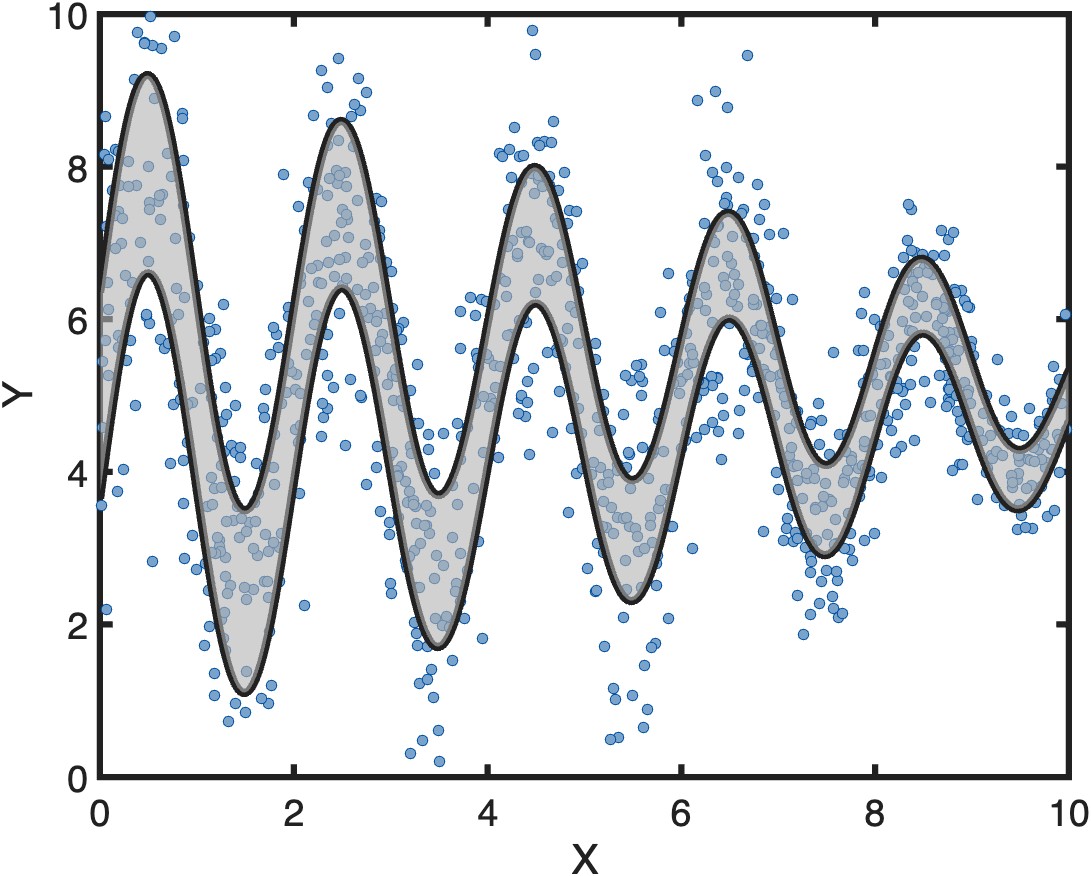}
        \caption{}
        \label{fig:d1_50mi}
    \end{subfigure}
    \hfill
    \begin{subfigure}{0.48\textwidth}
        \centering
        \includegraphics[width=\linewidth]{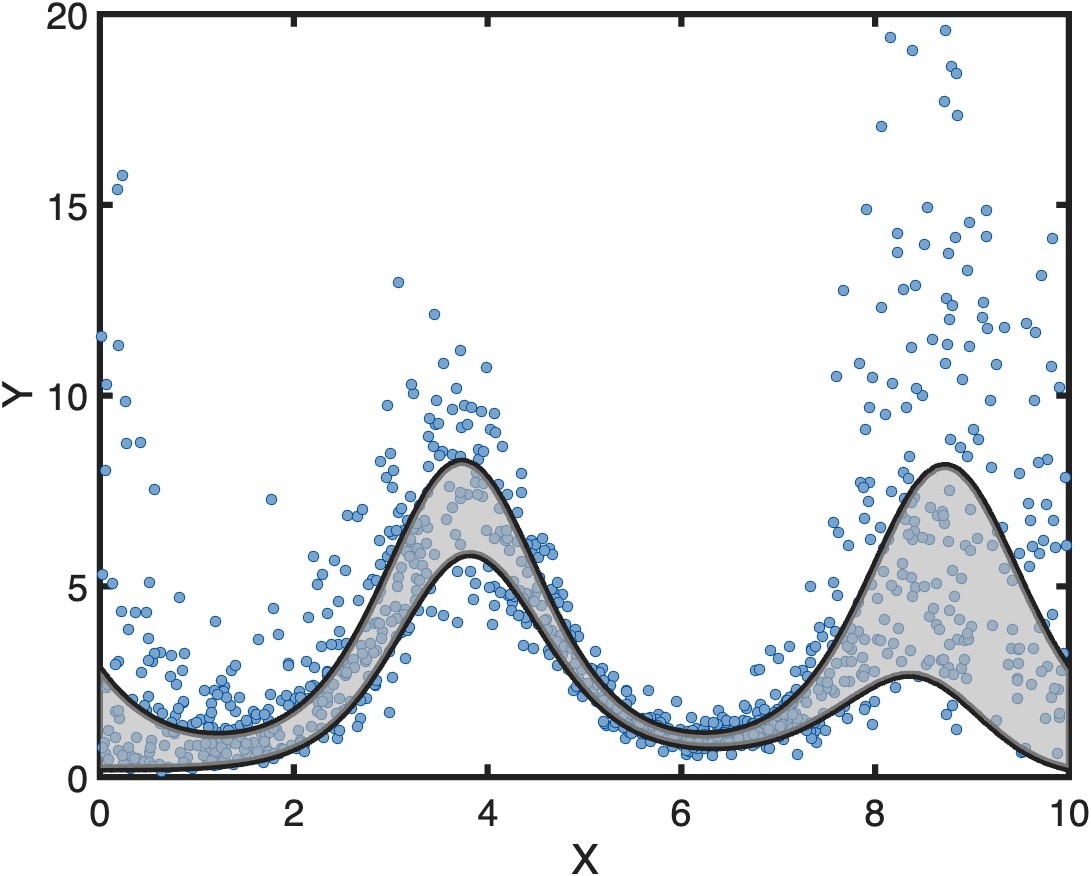}
        \caption{}
        \label{fig:d2_50mi}
    \end{subfigure}
    \caption{Visualization of the simulated data under (a) Distribution~1 and (b) Distribution~2. 
The band regions represent the true 50\% conditional MI theoretically derived from the known distributions, and the blue points denote 1,000 randomly generated data points from each distribution.}
    \label{fig:dis_example}
\end{figure}
Distribution~1 simulated a scenario in which the conditional distribution of \(Y\) changes rapidly with \(X\) to evaluate the performance of each method under sharp distributional variation.

\noindent\textbf{Distribution 1 (Normal \(Y\) with uniform \(X\))}: The conditional distribution of \(Y\) given \(X = x\) follows a symmetric normal distribution as given below.
\[
Y \mid X = x \sim \mathcal{N}(\mu_1(x), \sigma_1^2(x)),
\quad \text{with } X \sim \mathcal{U}(0, 10).
\]
The mean and SD functions are defined as
\[
\mu_1(x) \coloneqq (3 - 0.2x)\sin(\pi x) + 5, 
\quad 
\sigma_1(x) \coloneqq 2 - 0.15x, 
\quad \text{for } x \in (0, 10).
\]

Distribution~2 simulates a long-tailed distribution the tail length of which varies with \(X\) to evaluate each method under asymmetric and heavy-tailed conditions.

\noindent\textbf{Distribution 2 (Log-normal \(Y\) with uniform \(X\))}: The conditional distribution of \(Y\) given \(X = x\) is an asymmetric log-normal distribution,
\[
Y \mid X = x \sim \text{Log-normal}(\mu_2(x), \sigma_2^2(x)),
\quad \text{where } X \sim \mathcal{U}(0, 10).
\]
The mean and SD functions are defined as
\[
\mu_2(x) \coloneqq 1 - \sin(0.4\pi x), 
\quad 
\sigma_2(x) \coloneqq 0.04x^2 - 0.4x + 1.2, 
\quad \text{for } x \in (0, 10).
\]

For each dataset, we first sampled $n$ values of \(X\) values from a uniform distribution, , with $n$ chosen from four sample sizes.
The corresponding \(Y\) values were then generated according to the conditional distribution of \(Y \) given \(X = x\).
Figure~\ref{fig:dis_example} shows simulated datasets for both distributions.
As shown in the figure, the gray bands cover the areas in which the data were most concentrated, which indicates that the conditional MI captured the data concentration successfully.

\subsection{Parameter settings and evaluation metrics}
For the KDE method in Equation~(\ref{equ:kde_cmi}), we used the plug-in method \parencite{Sheather1991,Botev2010,Li2007} to select the bandwidth and apply a Gaussian kernel. The same bandwidth selection and kernel function were also used for the KDE in Equation~(\ref{equ:kde_x}). 

For both the bin-based MIR method and the proposed KDE-based MIR methods, spline knots were set to evenly divide the range from \(\xi_0 = 0\) to \(\xi_b = 10\) into \(b = 20\) subintervals, each with a length of \(\Delta_j = 0.5\) \((j = 1, 2, \ldots, b)\).
Increasing the spline degree beyond three provided minimal improvement but did lead to increased computational costs.
Therefore, we used cubic splines of degree \(d = 3\) and smoothness \(\rho = 2\).
The candidate values for \(\lambda\) are set to \(\lambda \in \{10^{-4}, 10^{-3}, 10^{-2}, 10^{-1}, 1\}\).

The performance of all methods was evaluated using three metrics, including RMSE, CP, and AIW, which respectively represent estimation accuracy, coverage reliability, and interval width.
RMSE was calculated at 101 points, with \(x_l = 0.1l\) \((l = 0, 1, \ldots, 100)\).
At each point, we compute the estimated upper bound \(\hat{y}_l^{\mathrm{up}}\) and lower bound \(\hat{y}_l^{\mathrm{low}}\), and compare them to the true values \(m_{\alpha,Y}^{\mathrm{up}}(x_l)\) and \(m_{\alpha,Y}^{\mathrm{low}}(x_l)\), which are theoretically calculated from the known distribution.
RMSE is defined as
\[
\mathrm{RMSE} \coloneqq 
\sqrt{\frac{1}{101} \sum_{l=1}^{101} \left(m_{\alpha,Y}^{\mathrm{up}}(x_l) - \hat{y}_l^{\mathrm{up}}\right)^2} 
+ 
\sqrt{\frac{1}{101} \sum_{l=1}^{101} \left(m_{\alpha,Y}^{\mathrm{low}}(x_l) - \hat{y}_l^{\mathrm{low}}\right)^2}.
\]

A separate test set containing 1{,}000 data points was used to compute CP and AIW.
CP represents the proportion of test samples contained within the estimated conditional MI and AIW denotes the average width of the estimated conditional MI evaluated at the test points.
\begin{table}[t]
\centering
\caption{Comparison of three methods across sample sizes under two distributions.
Each value represents the mean (SD) of RMSE, CP, and AIW over 50 repetitions, where CP is reported in percent.}
\label{tab:diff_n_exp_results}
\setlength{\tabcolsep}{4pt}
\small
\begin{tabular}{l c c c c c c}
\toprule
& \multicolumn{3}{c}{Distribution 1} & \multicolumn{3}{c}{Distribution 2} \\
\cmidrule(lr){2-4}\cmidrule(lr){5-7}
Method / \(n\) & RMSE & CP (\%) & AIW & RMSE & CP (\%) & AIW \\
\midrule
\multicolumn{7}{l}{\textit{Bin-based MIR}} \\
\(n=500\)  & 1.10 (0.18) & 43.9 (2.1) & 1.56 (0.09) & 1.20 (0.44) & 43.9 (2.2) & 1.86 (0.17) \\
\(n=1000\) & 0.89 (0.16) & 46.8 (1.8) & 1.65 (0.07) & 0.86 (0.21) & 46.8 (1.6) & 1.90 (0.09) \\
\(n=2000\) & 0.66 (0.14) & 48.4 (1.2) & 1.66 (0.06) & 0.63 (0.14) & 48.5 (1.1) & 1.95 (0.06) \\
\(n=3000\) & 0.61 (0.10) & 48.9 (1.0) & 1.67 (0.04) & 0.53 (0.11) & 49.3 (1.0) & 1.97 (0.06) \\
\midrule
\multicolumn{7}{l}{\textit{KDE-based MIR}} \\
\(n=500\)  & 1.04 (0.22) & 47.6 (2.0) & 1.70 (0.09) & 0.92 (0.25) & 48.2 (2.2) & 2.00 (0.16) \\
\(n=1000\) & 0.72 (0.18) & 49.4 (1.6) & 1.68 (0.07) & 0.72 (0.19) & 48.9 (1.3) & 2.00 (0.10) \\
\(n=2000\) & 0.63 (0.14) & 49.5 (1.1) & 1.70 (0.06) & 0.55 (0.15) & 49.8 (0.9) & 2.00 (0.07) \\
\(n=3000\) & 0.60 (0.14) & 49.9 (1.0) & 1.71 (0.04) & 0.51 (0.14) & 50.2 (0.9) & 2.02 (0.07) \\
\midrule
\multicolumn{7}{l}{\textit{KDE method}} \\
\(n=500\)  & 2.08 (0.39) & 56.2 (4.4) & 2.31 (0.21)  & 1.55 (0.28) & 52.8 (3.3) & 2.19 (0.16) \\
\(n=1000\) & 1.47 (0.22) & 56.8 (2.4) & 2.14 (0.13) & 1.10 (0.17) & 52.4 (2.0) & 2.09 (0.10) \\
\(n=2000\) & 1.05 (0.11) & 55.8 (3.6) & 1.99 (0.08) & 0.81 (0.13) & 47.9 (2.3) & 2.00 (0.07) \\
\(n=3000\) & 0.87 (0.10) & 48.9 (3.4) & 1.91 (0.06) & 0.68 (0.07) & 48.3 (2.6) & 1.91 (0.06) \\
\bottomrule
\end{tabular}
\end{table}
\begin{figure}[tb]
    \centering
    \begin{subfigure}{0.48\textwidth}
        \centering
        \includegraphics[width=\linewidth]{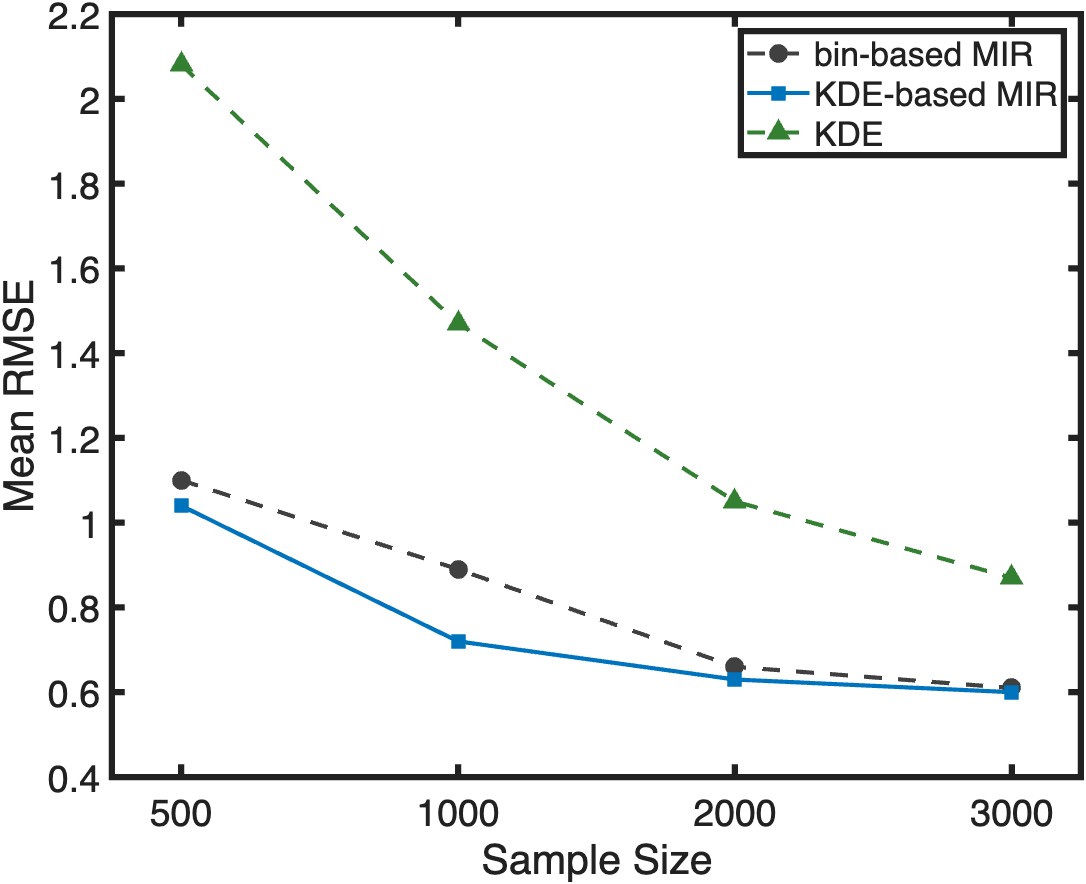}
        \caption{}
        \label{fig:plot_rmse_d1}
    \end{subfigure}
    \hfill
    \begin{subfigure}{0.48\textwidth}
        \centering
        \includegraphics[width=\linewidth]{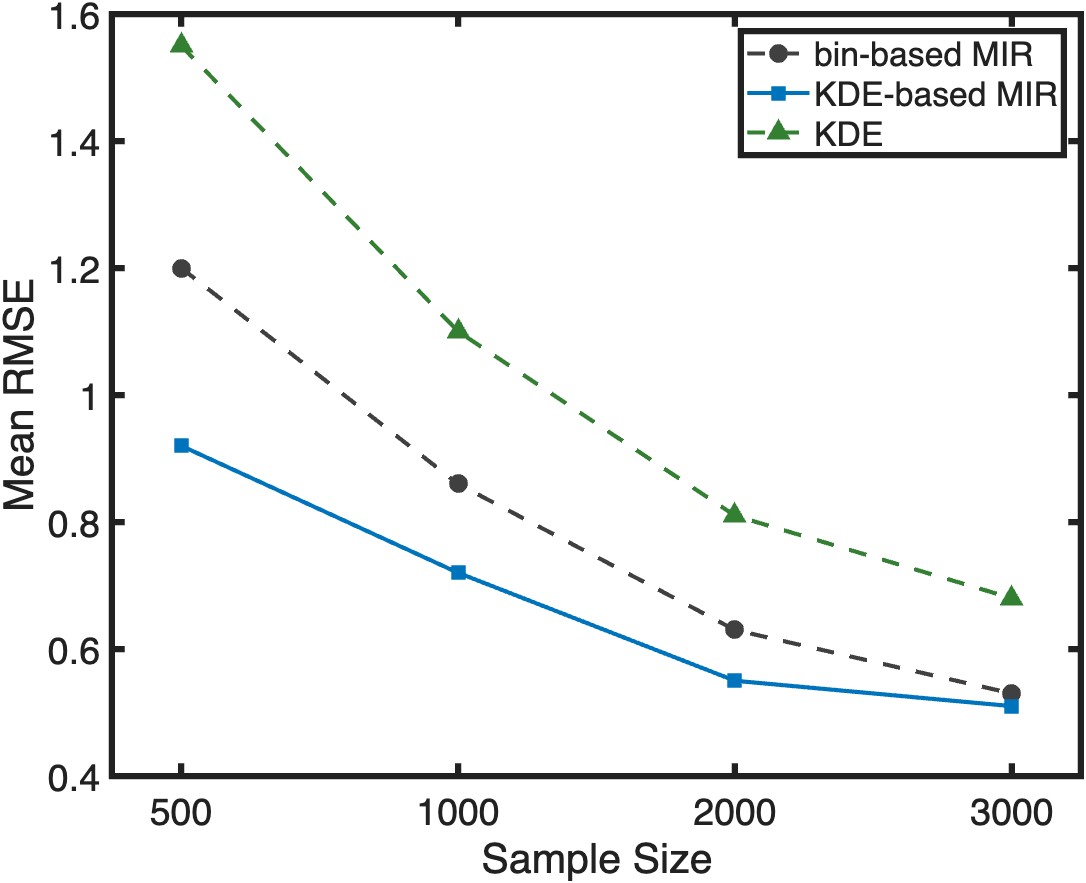}
        \caption{}
        \label{fig:plot_rmse_d2}
    \end{subfigure}
    \caption{RMSE versus sample size for the three methods under (a) Distribution~1 and (b) Distribution~2. The lines show the mean RMSE over 50 repetitions.}
    \label{fig:plot_rmse}
\end{figure}
\subsection{Effect of sample size on estimation accuracy and stability}
\label{sec:3-3}
First, we compare the performance of the three methods under different training sample sizes.
Table~\ref{tab:diff_n_exp_results} summarizes the results for two distributions.
Each value in the table is shown as the mean (SD) of RMSE, CP, and AIW over 50 repetitions.
For the bin-based MIR and KDE-based MIR methods, we report the results at the \(\lambda\) that yields the smallest RMSE for each sample size.
The full results for all candidate \(\lambda\) values are provided in Appendix~\ref{app:table}.
Figure~\ref{fig:plot_rmse} plots the mean RMSE of the three methods against the sample size.
The figure provides a direct visual comparison of estimation accuracy as the sample size increases.

\subsubsection{Results for Distribution 1}
For Distribution~1, all three methods become more accurate as the sample size increases.
Both the mean and SD of RMSE decrease with \(n\), which indicates that more data lead to more accurate and more stable estimation.
Across all sample sizes, the two MIR methods clearly outperform the KDE method in terms of RMSE.
This suggests that the MIR framework, which fits the MI bounds using the quantile loss with smoothing splines, provides a substantial accuracy gain over the tranditional KDE approach.

The KDE-based MIR is slightly more accurate than the bin-based MIR, with a more noticeable improvement around \(n=1000\), and the gap becomes smaller as \(n\) increases.
This is expected because the conditional distributions in Distribution~1 are symmetric, so the true quantile levels of the 50\% MI bounds are fixed at 0.25 and 0.75 and do not vary with \(x\), which limits the benefit of estimating quantile levels more flexibly.

In terms of CP, the KDE method shows CP values well above 50\% when the sample size is not large (\(n\le 2000\)).
This may be because the plug-in bandwidth tends to be relatively large, producing conservative (wider) intervals.
It is also less stable, as indicated by its larger SD compared with the two MIR methods.
Comparing the two MIR methods, when \(n\) is small (e.g., \(n=500\)), the bin-based MIR under-covers, with CP around 43.9\%, whereas the proposed KDE-based MIR is closer to the target, around 47.6\%.
As \(n\) increases, both methods approach the target coverage, but KDE-based MIR remains closer.
For example, at \(n=3000\) its CP is already about 49.9\%.

\subsubsection{Results for Distribution 2}

For Distribution~2, all three methods again become more accurate as the sample size increases.
The KDE method shows the same limitation as in Distribution~1.
Its RMSE is higher than those of the two MIR methods, which again indicates the benefit of the MIR framework.
However, because the conditional MI in Distribution~2 varies less sharply than in Distribution~1, the CP of the KDE method is closer to 50\%.

In terms of RMSE, Distribution~2 is more challenging because the conditional distribution is long-tailed and more complex.
This setting highlights the advantage of estimating quantile levels using KDE.
When the sample size is not large (\(n<3000\)), KDE-based MIR clearly outperforms bin-based MIR.
The difference becomes smaller as \(n\) increases, and the two methods perform similarly at \(n=3000\).

In terms of CP, both MIR methods move closer to the target 50\% coverage as \(n\) increases.
Bin-based MIR still under-covers when the sample size is small.
For example, at \(n=500\) its CP is only about 43.9\%.
KDE-based MIR remains closer to the target, with CP around 49.8\% and 50.2\% at \(n=2000\) and \(n=3000\), respectively.

Overall, these results highlight clear limitations of the KDE method.
It has lower accuracy, its coverage is less stable, and it tends to be overly conservative when the sample size is small.
When data are limited (\(n<3000\)), KDE-based MIR is more accurate than bin-based MIR.
The improvement is more pronounced when the conditional distribution is long-tailed and complex, as in Distribution~2.
In particular, bin-based MIR can suffer from under-coverage when the sample size is small.
In addition, KDE-based MIR yields coverage probabilities that are closer to the target level than those of bin-based MIR.

\subsection{Sensitivity to the smoothing parameter}
\label{sec:3-4}
Next, we study sensitivity to the smoothing parameter \(\lambda\) by comparing results across different \(\lambda\) values.
Table~\ref{tab:exp_results_kde_mir} reports the three metrics (RMSE, CP, and AIW) for all candidate \(\lambda\) values under two distributions with \(n=1000\).
Results for other sample sizes are provided in Appendix~\ref{app:table}.

\begin{table}[t]
\centering
\caption{Performance of the proposed method under two distributions with \(n=1000\) across different \(\lambda\) values.
Each value represents the mean (SD) of RMSE, CP, and AIW over 50 repetitions, where CP is reported in percent.}
\label{tab:exp_results_kde_mir}
\setlength{\tabcolsep}{4pt}
\small
\begin{tabular}{l c c c c c c}
\toprule
& \multicolumn{3}{c}{Distribution 1} & \multicolumn{3}{c}{Distribution 2} \\
\cmidrule(lr){2-4}\cmidrule(lr){5-7}
\(\lambda\) & RMSE & CP (\%) & AIW & RMSE & CP (\%) & AIW \\
\midrule
\(10^{-4}\) & 0.88 (0.18) & 48.7 (1.8) & 1.72 (0.07) & 0.97 (0.24) & 48.6 (1.5) & 2.02 (0.10) \\
\(10^{-3}\) & 0.81 (0.16) & 48.9 (1.9) & 1.71 (0.07) & 0.88 (0.23) & 48.7 (1.5) & 2.02 (0.10) \\
\(10^{-2}\) & 0.72 (0.19) & 49.4 (1.6) & 1.68 (0.07) & 0.72 (0.19) & 48.9 (1.3) & 2.00 (0.10) \\
\(10^{-1}\) & 1.80 (0.12) & 51.5 (1.4) & 1.98 (0.07) & 0.72 (0.17) & 49.4 (1.3) & 1.98 (0.10) \\
\(1\)       & 2.82 (0.06) & 52.0 (1.4) & 2.56 (0.08) & 1.77 (0.15) & 49.9 (1.3) & 2.12 (0.12) \\
\bottomrule
\end{tabular}
\end{table}

From Table~\ref{tab:exp_results_kde_mir}, we can see several clear patterns as \(\lambda\) varies.
For both distributions, CP increases as \(\lambda\) increases, and its variability decreases (smaller SD).
This is because a stronger roughness penalty produces smoother and more conservative fitted bounds.
In terms of RMSE, we observe a typical U-shaped trend in that small $\lambda$ values lead to overfitting, while large $\lambda$ values resulted in over-smoothing.

For Distribution~1, \(\lambda=10^{-2}\) performs best.
It yields the smallest RMSE, a CP close to 50\%, and the shortest AIW.
For Distribution~2, \(\lambda=10^{-1}\) is preferred.
Although its RMSE is very similar to that of \(\lambda=10^{-2}\), it shows smaller variability, a CP closer to 50\%, and a slightly shorter AIW.

\subsection{Comparison of computation time}
\label{sec:3-5}
Finally, we compare the computation times of the proposed method across different sample sizes with \(\lambda = 10^{-2}\).
We ran all experiments in MATLAB R2023b (64-bit) on macOS Tahoe (Apple M1 Pro SoC, 16 GB RAM).

In Step 1, we estimate the conditional distribution by Equation~\eqref{equ:kde_cdf}.
This step can be slow when $n$ is large.
We use all observations when $n\leq 1000$, and randomly sample 1000 observations when $n> 1000$ to estimate the conditional distribution.
This substantially reduces runtime for large $n$, at the cost of a trade-off between accuracy and computation time.
In Step 2, we fix the number of ADMM iterations at 1000.

Table~\ref{tab:time_stage_dist_n} reports the average computation time over 50 repetitions, including Step 1, Step 2, and the total time, for two distributions and four sample sizes.
Overall, the runtime increases with the sample size, while the effect of the distribution is small.
Step~1 accounts for a larger share of the total computation time.

\begin{table}[t]
\centering
\caption{Computation time (s) of the proposed method by stage under two distributions and four training sample sizes. Each value is the average over 50 repetitions.}
\label{tab:time_stage_dist_n}
\small
\setlength{\tabcolsep}{4pt}
\begin{tabular}{l *{4}{c} *{4}{c}}
\toprule
& \multicolumn{4}{c}{Distribution 1} & \multicolumn{4}{c}{Distribution 2} \\
\cmidrule(lr){2-5}\cmidrule(lr){6-9}
Stage
& \(n=500\) & \(n=1000\) & \(n=2000\) & \(n=3000\)
& \(n=500\) & \(n=1000\) & \(n=2000\) & \(n=3000\) \\
\midrule
Step 1 & 3.0 & 24.7 & 49.8 & 73.4 & 3.1 & 24.2 & 48.6 & 76.3 \\
Step 2 & 1.7 & 3.4  & 6.8  & 9.5  & 1.8 & 3.5  & 6.5  & 9.7 \\
Total  & 4.7 & 28.1 & 56.6 & 82.9 & 4.9 & 27.7 & 55.1 & 86.0 \\
\bottomrule
\end{tabular}
\end{table}

From $n=500$ to $n=1000$, the sample size doubles, but the Step~1 time increases by about eight times. 
This is expected when Step~1 uses all observations. 
For a sample size $n$, Equation~\eqref{equ:kde_cdf} is computed at $n$ points $\{x_i\}_{i=1}^n$ using all $n$ observations, resulting in $O(n^2)$ computations.
In addition, finding the MI is performed at each $x_i$ and may require up to $O(n^2)$ scans in the worst case, yielding $O(n^3)$ scans in total.
Both costs grow quickly with $n$, so Step~1 can become a bottleneck.
For this reason, we set a threshold in Step~1 and use at most 1000 observations to estimate the conditional distribution. 
This may reduce accuracy, but it greatly reduces computation time.

After this threshold is applied, the runtime grows roughly linearly with \(n\), as seen for \(n=1000, 2000,\) and \(3000\).
Step~2 also increases approximately linearly with \(n\) when the number of ADMM iterations is fixed, and it accounts for only a small portion of the total runtime because we solve the convex optimization problem in Step~2 using ADMM.

In practice, users can choose the Step~1 threshold based on their hardware.
A larger threshold may improve accuracy but increases computation time, while a smaller threshold saves time at the cost of accuracy.
For large datasets, one can first run the method on a smaller subsample and use the measured time to predict the runtime for the full sample.

\section{Analysis of real neonatal hormone data}
\subsection{Data description and preprocessing}
Melatonin and cortisol are key hormones that regulate circadian rhythms in the human body. 
Melatonin, produced by the pineal gland, helps control sleep-wake cycles, whereas cortisol, secreted by the adrenal cortex, plays a role in stress response and metabolism. 
Both hormones exhibit clear circadian patterns \parencite{Selmaoui2003,Claustrat2005,Fries2009,Skubic2025}. 
However, these rhythms differ between neonates and adults. 
Understanding hormonal rhythms in newborns may help improve their sleep patterns \parencite{MunozHoyos1993,DeWeerth2003,Iwata2013,Tuladhar2021,Wong2022}.

Previous studies on neonatal cortisol rhythms \parencite{Iwata2013} have, for example, grouped several days of measurements into four six-hour intervals to test a hypothesis. 
However, this grouping may obscure daily changes, as hormonal patterns can vary from day to day.
This approach primarily tests whether the mean level differs across the predefined intervals, while it provides limited information about how hormone levels change continuously over time.
A continuous characterization of temporal changes may therefore be informative for understanding neonatal hormonal rhythms.
Moreover, the data exhibited skewed and long-tailed distributions, and substantial individual differences among neonates made it difficult to represent circadian rhythms with a single, uniform pattern. 

To address these issues, we applied the proposed nonlinear MIR method to estimate the 50\% conditional MI. 
This interval represents the most concentrated interval of the data and provides a robust estimate of the neonatal hormone rhythms evident in the data. 
The dataset was provided by the Kurume University School of Medicine and included hormone measurements from 33 neonates during their first 10 days after birth for a total of 1,652 observations. 
Data were collected between December 2012 and January 2015. 
Cortisol and 6-sulfatoxymelatonin (the main metabolite of melatonin, hereafter referred to as melatonin) levels were measured using urine samples at multiple time points each day, approximately every three hours from 1:00~AM to 10:00~PM.

Our goal was to characterize, separately, how neonatal cortisol and melatonin levels change continuously over time, from 6:00~AM on postnatal day 2 to 6:00~PM on postnatal day 10. By doing so, we aim to explore the development of neonatal circadian rhythms.

During preprocessing, the original time variable \(X\) (recorded in hour format, e.g., 00:00) was converted into a continuous variable. 
We set \(X = 0\) at 00:00 on the day of birth and increased \(X\) by 0.1 per hour to represent time continuously over the 10 days. 
To minimize the influence of immediate postnatal effects, the analysis began at 6:00~AM on the second day and ended at 6:00~PM on the tenth day \((X \in [3, 23.4])\). 
Here, \(X = 3\) corresponds to 6:00~AM on the second day. 
In this study, we focused on analyzing the more stable physiological conditions.

For cortisol, the time \(X\) and cortisol concentration \(Y_1\) (nmol/L) formed the bivariate pair \((X, Y_1)\), with \(n = 1311\) observations \(\{(x_i, y_{1,i})\}_{i=1}^{1311}\). 
For melatonin, \((X, Y_2)\) were defined similarly, with \(n = 1297\) observations \(\{(x_i, y_{2,i})\}_{i=1}^{1297}\).

\subsection{Parameter settings and analysis results}
\subsubsection{Parameter settings}
\begin{figure}[tbp]
    \centering
    \begin{subfigure}[t]{0.48\textwidth}
        \centering
        \includegraphics[width=\linewidth]{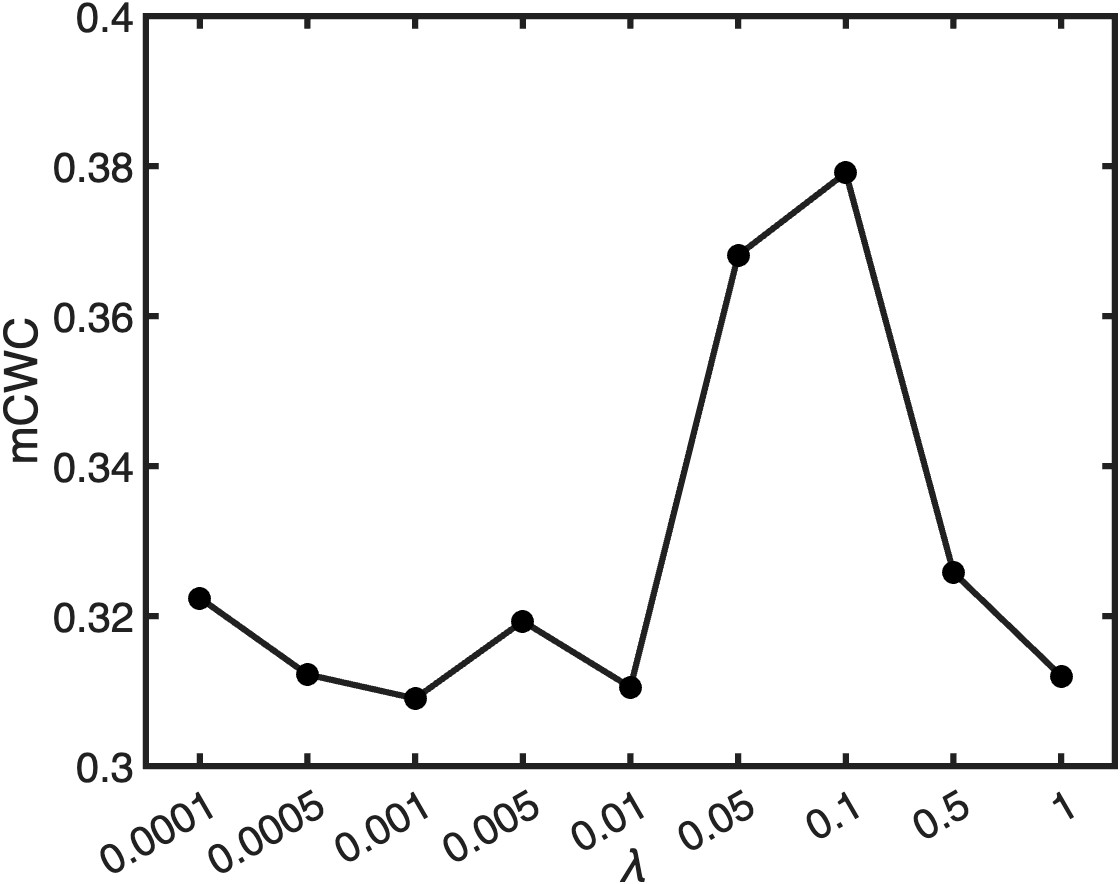}
        \caption{}
        \label{fig:cor}
    \end{subfigure}
    \hfill
    \begin{subfigure}[t]{0.48\textwidth}
        \centering
        \includegraphics[width=\linewidth]{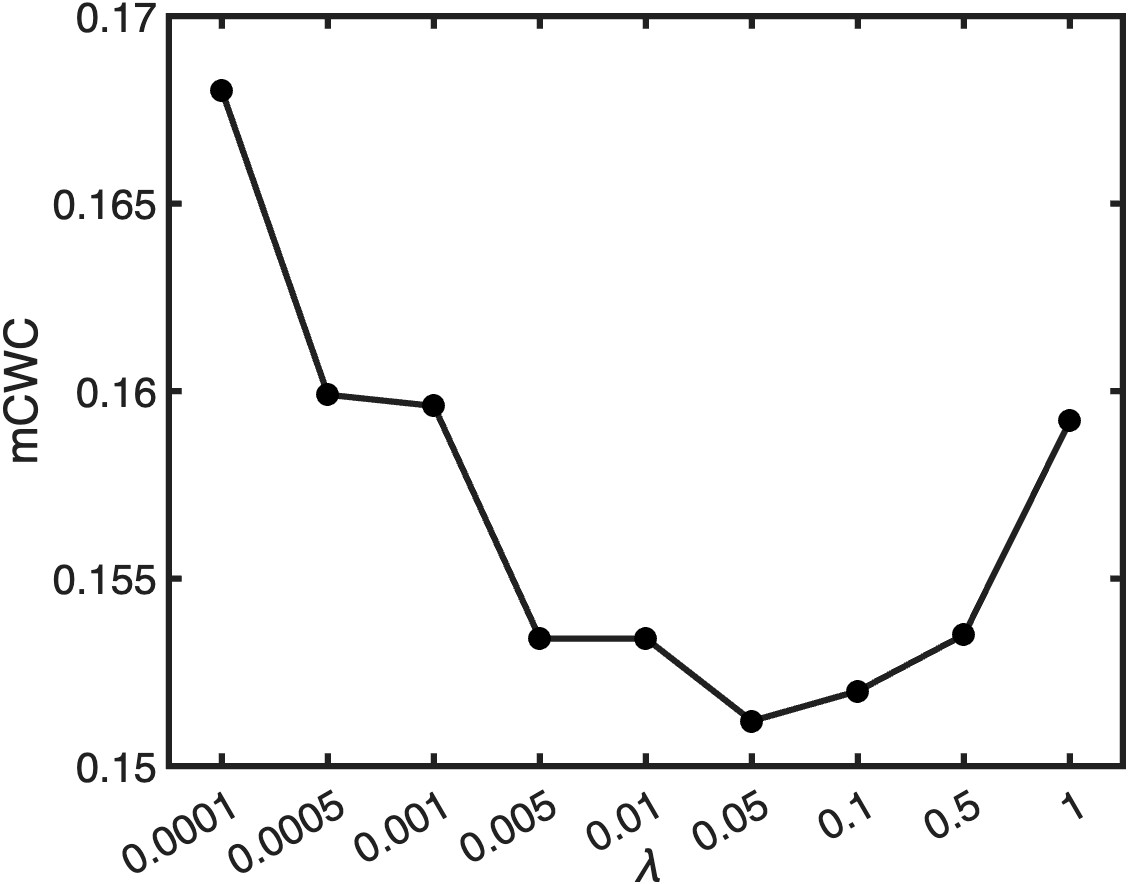}
        \caption{}
        \label{fig:mel}
    \end{subfigure}
    \caption{Effects of different $\lambda$ values on average mCWC for cortisol (a) and melatonin (b). 
  The minimum mCWC was achieved at $\lambda = 0.001$ for cortisol and $\lambda = 0.05$ for melatonin.}
    \label{fig:mcwc_lambda}
\end{figure}

For the proposed nonlinear MIR method in Equation~(\ref{equ:joint_opt}), spline knots were set from \(\xi_0 = 3\) and \(\xi_{20} = 23.4\) with equal subinterval width \(\Delta_j = 1.02\) \((j = 1, 2, \ldots, 20 =: b)\). 
We used cubic splines with degree \(d = 3\) and smoothness \(\rho = 2\).
The smoothing parameter \(\lambda\) was selected from the set \(\{0.0001, 0.0005, 0.001, 0.005, \dots, 0.5, 1\}\) using 5-fold cross-validation. 
The average mCWC over the five folds was computed for each \(\lambda\). 
Figure~\ref{fig:mcwc_lambda} shows the results, with optimal \(\lambda\) values of 0.001 for cortisol and 0.05 for melatonin.

\subsubsection{Detection of hormonal rhythms}
To define a hormonal rhythm, a local maximum was identified if the hormone level at the time was higher than all levels in the preceding and following 24-hour periods, with local minima occurring in both. 
The peak was also required to exceed both minima by a specified ratio. 
The interval between the first and second minima defined a rhythmic cycle, with the peak representing the maximum level of the cycle. 

We classified the rhythms into three levels based on the ratio of the peak to the minimum calculated using the midpoint of the MI. 
Given the relatively small hormonal fluctuations observed in neonates, the threshold values were set lower. 
A ratio of 1.5 or higher indicates a significant rhythm, a ratio between 1.25 and 1.5 indicates a mild rhythm, and a ratio below 1.25 indicates either no rhythm or minor fluctuations.

\begin{figure}[tbp]
    \centering
    \includegraphics[width=1\linewidth]{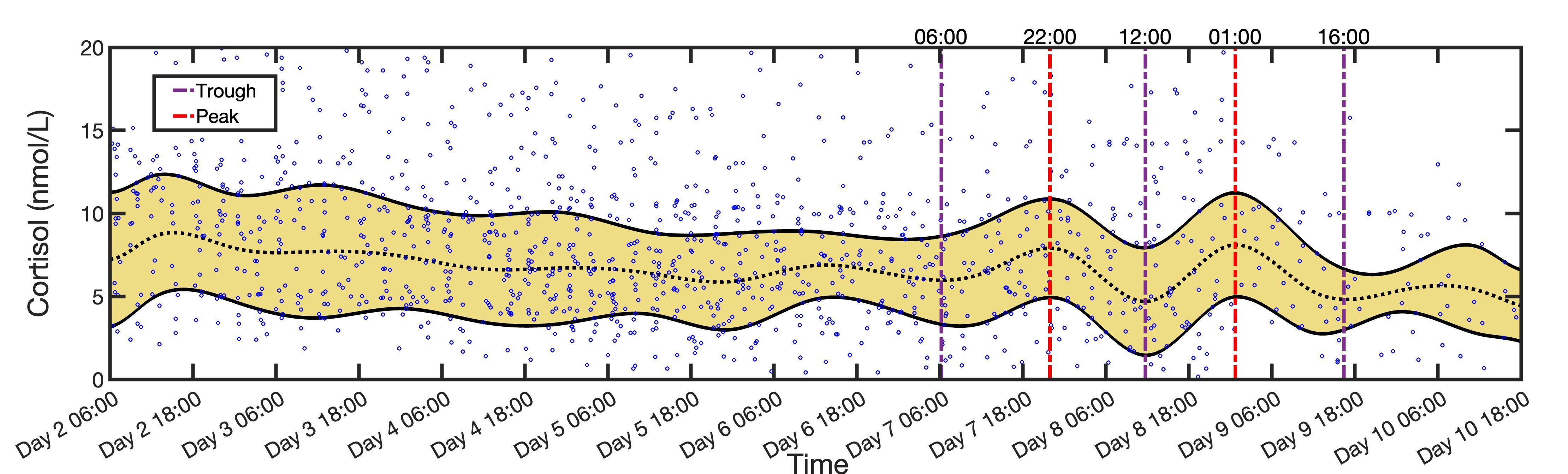}
    \caption{MIR results for cortisol levels. The yellow band represents the 50\% MI, and the dotted line represents the midpoint (the average of the upper and lower bounds). Purple dashed-dotted lines mark circadian troughs, and red lines mark peaks.}
    \label{fig:cor_mi}
\end{figure}
\begin{figure}[tbp]
    \centering
    \includegraphics[width=1\linewidth]{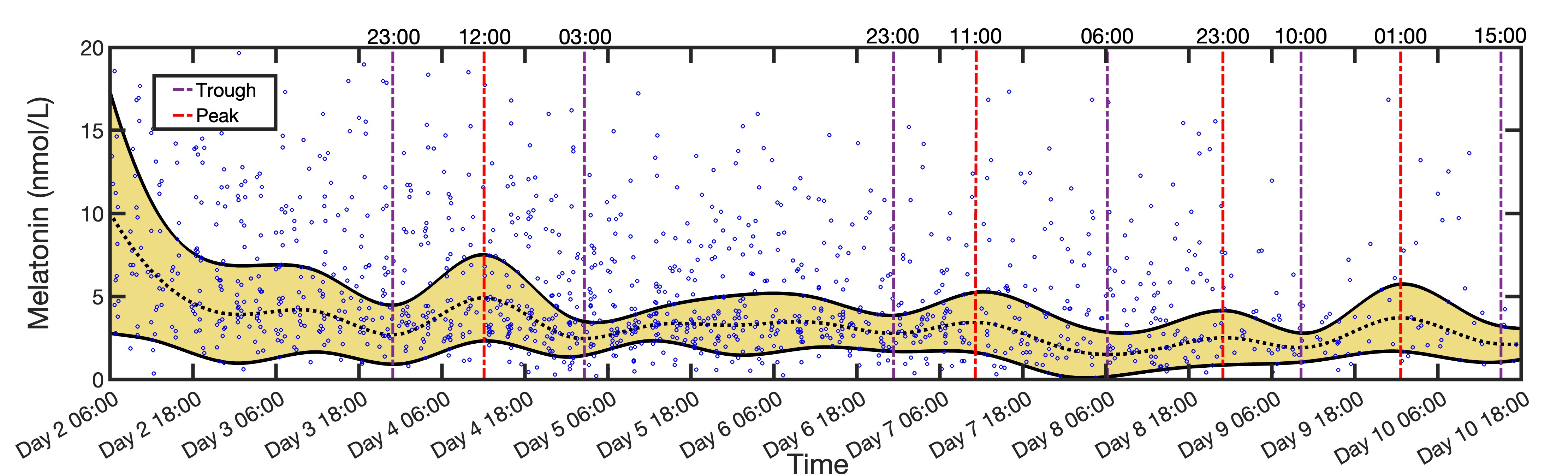}
    \caption{MIR results for melatonin levels. The yellow band represents the 50\% MI, and the dotted line represents the midpoint (the average of the upper and lower bounds). Purple dashed-dotted lines mark circadian troughs, and red lines mark peaks.}
    \label{fig:mel_mi}
\end{figure}

\subsubsection{Results for cortisol}
Figure~\ref{fig:cor_mi} illustrates the temporal pattern of cortisol levels estimated using MIR from 6:00~AM on Day~2 to 6:00~PM on Day~10.
No clear rhythm was observed before 6:00~AM on Day~7. 
The first rhythmic interval was detected from 6:00~AM on Day~7 to 12:00~PM on Day~8, with a peak at 10:00~PM on Day~7 lasting for approximately 30 hours. 
The 50\% MIs at 6:00~AM and 10:00~PM on Day~7, and 12:00~PM on Day~8 were \([3.31, 8.62]\), \([4.92, 10.86]\), and \([1.45, 7.93]\)~nmol/L, respectively. 
The midpoint ratios were 1.32 and 1.68, indicating a mild rhythm. 

The second rhythm appeared from 12:00~PM on Day~8 to 4:00~PM on Day~9, peaking at 1:00~AM on Day~9 and lasting approximately 28 hours. 
The 50\% MIs at the trough, peak, and following trough were \([1.45, 7.93]\), \([4.97, 11.23]\), and \([3.00, 6.65]\)~nmol/L, with midpoint ratios of 1.72 and 1.68, indicating a significant rhythm. 
A possible third rhythm emerged after 4:00~PM on Day~9, peaking at 7:00~AM on Day~10. 
However, the midpoint ratio was only 1.17, which suggests minor fluctuations rather than a clear rhythm. 

\subsubsection{Results for melatonin}
Figure~\ref{fig:mel_mi} shows the temporal pattern of melatonin levels estimated using MIR from 6:00~AM on Day~2 to 6:00~PM on Day~10. 
Melatonin levels dropped sharply from \([2.78, 17.35]\)~nmol/L at 6:00~AM on Day~2 to \([0.91, 4.47]\)~nmol/L at 11:00~PM on Day~3. 
The first rhythmic interval appeared from 11:00~PM on Day~3 to 3:00~AM on Day~5, peaking at 12:00~PM on Day~4 and lasting for approximately 28 hours. 
The 50\% MIs at the trough, peak, and following trough were \([0.91, 4.47]\), \([2.31, 7.50]\), and \([1.45, 3.47]\)~nmol/L, with midpoint ratios of 1.82 and 1.99, indicating a significant rhythm. 

No clear rhythm was observed until 11:00~PM on Day~6. 
The second rhythmic interval spanned from 11:00~PM on Day~6 to 6:00~AM on Day~8, peaking at 11:00~AM on Day~7 (approximately 31 hours). 
The 50\% MIs at the trough, peak, and following trough were \([1.63, 3.81]\), \([1.60, 5.25]\), and \([0.15, 2.83]\)~nmol/L, with midpoint ratios of 1.26 and 2.30, indicating a mild rhythm. 

The third rhythmic interval occurred from 6:00~AM on Day~8 to 10:00~AM on Day~9, peaking at 11:00~PM on Day~8 (approximately 28 hours). 
The 50\% MIs at the trough, peak, and following trough were \([0.15, 2.83]\), \([0.87, 4.14]\), and \([1.04, 2.78]\)~nmol/L, with midpoint ratios of 1.68 and 1.31, indicating a mild rhythm. 

The fourth rhythm was observed from 10:00~AM on Day~9 to 3:00~PM on Day~10, peaking at 1:00~AM on Day~10 (approximately 29 hours). 
The 50\% MIs at the trough, peak, and following trough were \([1.04, 2.78]\), \([1.68, 5.74]\), and \([1.05, 3.19]\)~nmol/L, with midpoint ratios of 1.94 and 1.75, indicating a significant rhythm.

\subsection{Discussion}
\subsubsection{Overall interpretation}
Given the large differences between individuals among neonates, we did not aim to identify specific rhythms for each neonate or a shared rhythm across all individuals. 
Instead, MIR was used to estimate the 50\% MIs of cortisol and melatonin levels, which are the most frequently observed hormones. 
For the unimodal distributions assumed in this study, the 50\% MI captured the core region around the peak, covering at least half of the neonates. 
This approach provides valuable insight into the general hormonal trends during the first ten days after birth. 

As shown in Figures~\ref{fig:cor_mi} and~\ref{fig:mel_mi}, the 50\% MI levels of both hormones were significantly lower than those reported in adults. 
This is likely due to the immature endocrine system of neonates. 
Additionally, clear circadian rhythms have not yet been established, and the day-night patterns of cortisol and melatonin remain incomplete and irregular compared to those in adults. 

\subsubsection{Cortisol rhythms}
As noted in Figure~\ref{fig:cor_mi}, no clear cortisol rhythm was observed before Day~7, likely because the newborns’ biological clocks were still developing after birth. 
Around Day~7, a weak rhythm appeared, which peaked at 10:00~PM and lasted for approximately 30~hours. 
This pattern differs from the typical adult 24-hour cycle, in which cortisol levels peak in the morning. 
This difference in timing may be attributed to the longer sleep duration in neonates. 

The second rhythm lasted 28~hours and peaked approximately three hours later than the first, suggesting a gradual shift toward a more regular pattern. 
No distinct third rhythm was observed; however, the overall trend indicates that as neonatal rhythms mature, cortisol peaks may gradually shift toward the morning to resemble the adult pattern. 
The absence of a clear third rhythm could be due to limited data, and further observations would be required to confirm this trend. 

\subsubsection{Melatonin rhythms}
As shown in Figure~\ref{fig:mel_mi}, from 6:00~AM on Day~2 to 11:00~PM on Day~3, melatonin levels decreased rapidly, and the 50\% MI range narrowed. 
This suggests that neonates initially receive a large amount of maternal melatonin, which declines sharply after birth due to their immature secretory system. 
The wide initial 50\% MI reflected individual differences in maternal melatonin transfer. 
A small fluctuation, with a peak on the morning of Day~3, was observed during this decline. 

A distinct rhythm began at 11:00~PM on Day~3, peaking at noon, with a period of approximately 28~hours. 
This first rhythm is likely influenced by residual maternal melatonin, resulting in a less organized pattern than that observed in adults. 
No clear rhythm was observed until 11:00~PM on Day~6, likely because maternal melatonin had been depleted and the neonates were not yet able to produce sufficient amounts of melatonin on their own. 

The second rhythm began at 11:00~PM on Day~6 and peaked at 11:00~AM, with a period of approximately 31~hours. 
Both the period and peak timing differ from adult patterns, which reflects neonatal sleep differences and endocrine immaturity. 
The third and fourth rhythms peaked at 11:00~PM and 1:00~AM, with periods of approximately 28 and 29~hours, respectively. 
From the third rhythm onward, the neonatal patterns gradually began to resemble adult rhythms, although the periods remained longer. 
The fourth rhythm was stronger than the third, which suggests the maturation of melatonin production and further development of the biological clock.

\section{Discussion}
In this study, we have reformulated the MIR framework using KDE for more effective analysis and visualization of bivariate data.
The proposed method estimates the data concentration interval in two steps.
First, the quantile levels corresponding to the conditional MI bounds are estimated using KDE.
Second, smoothing splines with a quantile loss function are applied to simultaneously estimate the upper and lower bounds.
The problem is then reformulated as a convex optimization problem, which is solved efficiently using the ADMM algorithm.
We implemented mCWC in our method to select parameters.

The results of numerical experiments demonstrated that the proposed method achieved higher accuracy and greater stability than both the traditional KDE and bin-based MIR methods, particularly for complex distributions.
The analysis of real data on neonatal hormone levels further confirmed its robustness in handling heavy-tailed distributions and its capability to visualize data concentration intervals effectively. 
By focusing on the MI, which represents the region in which hormone levels were most concentrated, we obtained a new perspective for analyzing hormonal data and revealed some potential underlying rhythmic patterns in neonatal hormone secretion.

We also applied MIR to time-dependent hormone data, and the results demonstrate its potential to capture temporal dynamics in data concentration. 
Building on this capability, we plan to extend the framework to spatial data analysis to enable the estimation and visualization of spatially varying data concentration intervals and spatial changes in their degree of concentration.

This extension involves two covariates.
We would therefore use KDE with bivariate covariates to estimate the quantile levels corresponding to the conditional MI bounds.
We would also replace the current spline curves with spline surfaces for the upper and lower bounds.
This increases the number of spline coefficients to be estimated, but the resulting optimization problem can still be solved efficiently using our ADMM-based algorithm.
In addition, the non-crossing constraint must be extended from curves to surfaces to ensure that the upper surface stays above the lower surface over the spatial domain.

Extending the framework to higher-dimensional covariates is more challenging. Multivariate KDE and spline regression both become less effective as the covariate dimension increases, and such settings may require alternative modeling approaches.

\section*{Acknowledgments}
This study was supported by JST SPRING (grant number JPMJSP2114).
\printbibliography
\appendix
\section{Additional Tables}
\label{app:table}
\centering
\vspace*{-6mm}
\begin{sideways}
\begin{minipage}{\textheight}
\captionsetup{width=.9\textheight}
\captionof{table}{Results for Distribution~1 across sample sizes and smoothing parameters.
Each value represents the mean (SD) of RMSE,\\ CP, and AIW over 50 repetitions, where CP is reported in percent.}
\label{tab:exp_results_n_1}
\setlength{\tabcolsep}{4pt}
\begin{tabular}{l *{4}{ccc}}
\toprule
& \multicolumn{3}{c}{$n=500$}
& \multicolumn{3}{c}{$n=1000$}
& \multicolumn{3}{c}{$n=2000$}
& \multicolumn{3}{c}{$n=3000$} \\
\cmidrule(lr){2-4}\cmidrule(lr){5-7}\cmidrule(lr){8-10}\cmidrule(lr){11-13}
$\lambda$ & RMSE & CP (\%) & AIW & RMSE & CP (\%) & AIW & RMSE & CP (\%) & AIW & RMSE & CP (\%) & AIW \\
\midrule
\multicolumn{13}{l}{\textit{Bin-based MIR}}\\
$10^{-4}$ & 1.16 (0.18) & 43.6 (2.1) & 1.56 (0.08) & 0.92 (0.15) & 46.6 (1.5) & 1.64 (0.07) & 0.70 (0.11)& 48.3 (1.2) & 1.67 (0.06) & 0.63 (0.10) & 48.9 (1.0) & 1.67 (0.04)  \\
$10^{-3}$ & 1.10 (0.18) & 43.9 (2.1) & 1.57 (0.09) & 0.89 (0.16) & 46.8 (1.8) & 1.65 (0.07) & 0.67 (0.11) & 48.4 (1.2) & 1.67 (0.06) & 0.61 (0.10) & 48.9 (1.0) & 1.67 (0.04) \\
$10^{-2}$ & 1.26 (0.23) & 44.2 (1.9) & 1.60 (0.09) & 0.95 (0.20) & 47.3 (1.6) & 1.64 (0.06) & 0.66 (0.14) & 48.4 (1.2) & 1.66 (0.06) & 0.61 (0.13) & 48.9 (0.9) & 1.67 (0.04) \\
$10^{-1}$ & 2.44 (0.15) & 45.8 (2.3) & 2.14 (0.11) & 2.01 (0.15) & 49.8 (1.5) & 2.01 (0.08) & 1.45 (0.14) & 48.4 (0.8) & 1.83 (0.05) & 1.17 (0.13) & 49.2 (0.8) & 1.77 (0.04) \\
$1$       & 2.98 (0.09) & 46.0 (2.6) & 2.47 (0.15) & 2.87 (0.07) & 50.5 (1.3) & 2.51 (0.08) & 2.69 (0.06) & 48.7 (1.0) & 2.44 (0.05) & 2.55 (0.06) & 49.2 (0.8) & 2.38 (0.05) \\
\midrule
\multicolumn{13}{l}{\textit{KDE-based MIR}}\\
$10^{-4}$ & 1.15 (0.22) & 47.7 (2.1) & 1.73 (0.08) & 0.88 (0.18) & 48.7 (1.8) & 1.72 (0.07) & 0.76 (0.15) & 49.4 (1.1) & 1.71 (0.05) & 0.71 (0.13) & 49.7 (1.1) & 1.72 (0.04) \\
$10^{-3}$ & 1.04 (0.22) & 47.6 (2.0) & 1.70 (0.09) & 0.81 (0.16) & 48.8 (1.9) & 1.71 (0.07) & 0.63 (0.14) & 49 (1.0) & 1.70 (0.06) & 0.71 (0.15) & 49.8 (1.0) & 1.72 (0.04) \\
$10^{-2}$ & 1.09 (0.27) & 47.8 (2.0) & 1.70 (0.10) & 0.72 (0.18) & 49.4 (1.6) & 1.68 (0.07) & 0.63 (0.14) & 49.5 (1.1) & 1.70 (0.06) & 0.60 (0.14) & 49.9 (1.0) & 1.71 (0.04) \\
$10^{-1}$ & 2.35 (0.13) & 48.0 (2.3) & 2.19 (0.11) & 1.80 (0.12) & 51.5 (1.4) & 1.98 (0.07) & 1.29 (0.11) & 49.4 (0.8) & 1.81 (0.05) & 1.05 (0.19) & 50.0 (0.9) & 1.77 (0.05) \\
$1$       & 2.98 (0.08) & 48.4 (3.0) & 2.60 (0.16) & 2.82 (0.06) & 52.0 (1.4) & 2.56 (0.08) & 2.64 (0.04) & 49.4 (1.1) & 2.45 (0.06) & 2.48 (0.04) & 49.8 (0.9) & 2.36 (0.05) \\
\bottomrule
\end{tabular}
\end{minipage}
\end{sideways}

\begin{sideways}
\begin{minipage}{\textheight}
\captionsetup{width=.9\textheight}
\captionof{table}{Results for Distribution~2 across sample sizes and smoothing parameters.
Each value represents the mean (SD) of RMSE, \\CP, and AIW over 50 repetitions, where CP is reported in percent.}
\label{tab:exp_results_n_2}
\setlength{\tabcolsep}{4pt}
\begin{tabular}{l *{4}{ccc}}
\toprule
& \multicolumn{3}{c}{$n=500$}
& \multicolumn{3}{c}{$n=1000$}
& \multicolumn{3}{c}{$n=2000$}
& \multicolumn{3}{c}{$n=3000$} \\
\cmidrule(lr){2-4}\cmidrule(lr){5-7}\cmidrule(lr){8-10}\cmidrule(lr){11-13}
$\lambda$ & RMSE & CP (\%) & AIW & RMSE & CP (\%) & AIW & RMSE & CP (\%) & AIW & RMSE & CP (\%) & AIW \\
\midrule
\multicolumn{13}{l}{\textit{Bin-based MIR}}\\
$10^{-4}$ & 1.65 (0.50) & 43.3 (2.3) & 1.90 (0.16) & 1.16 (0.29) & 46.6 (1.8) & 1.94 (0.1) & 0.85 (0.25) & 48.1 (1.1) & 1.97 (0.07) & 0.65 (0.14) & 49.1 (1.0) & 1.99 (0.06) \\
$10^{-3}$ & 1.47 (0.49) & 43.6 (2.3) & 1.90 (0.16) & 1.08 (0.28) & 46.7 (1.8) & 1.94 (0.10) & 0.79 (0.23) & 48.1 (1.1) & 1.97 (0.07) & 0.63 (0.14) & 49.1 (1.0) & 1.99 (0.06) \\
$10^{-2}$ & 1.20 (0.44) & 43.9 (2.2) & 1.86 (0.17) & 0.92 (0.24) & 46.8 (1.8) & 1.93 (0.10) & 0.69 (0.22) & 48.2 (1.2) & 1.96 (0.07) & 0.57 (0.13) & 49.1 (1.0) & 1.98 (0.06) \\
$10^{-1}$ & 1.25 (0.39) & 44.9 (2.4) & 1.88 (0.16) & 0.86 (0.21) & 46.8 (1.6) & 1.90 (0.09)  & 0.63 (0.14) & 48.5 (1.1) & 1.95 (0.06) & 0.53 (0.11) & 49.3 (1.0) & 1.97 (0.06) \\
$1$       & 2.30 (0.29) & 46.3 (1.8) & 2.11 (0.15) & 1.76 (0.19) & 46.6 (1.5) & 2.00 (1.11) & 1.28 (0.11) & 48.9 (1.0) & 1.99 (0.07) & 1.04 (0.14) & 49.4 (0.9) & 1.98 (0.05) \\
\midrule
\multicolumn{13}{l}{\textit{KDE-based MIR}}\\
$10^{-4}$ & 1.31 (0.32) & 47.1 (2.3) & 2.03 (0.17) & 0.97 (0.24) & 48.6 (1.5) & 2.02 (0.10) & 0.78 (0.24) & 49.5 (0.9) & 2.03 (0.07) & 0.68 (0.20) & 49.9 (0.9) & 2.04 (0.07) \\
$10^{-3}$ & 1.13 (0.28) & 47.5 (2.1) & 2.02 (0.16) & 0.88 (0.23) & 48.7 (1.5) & 2.02 (0.10) & 0.73 (0.23) & 49.6 (1.0) & 2.03 (0.08) & 0.66 (0.18) & 50.0 (1.0) & 2.04 (0.06) \\
$10^{-2}$ & 0.92 (0.25) & 48.2 (2.2) & 2.00 (0.16) & 0.72 (0.19) & 48.9 (1.3) & 2.00 (0.10) & 0.61 (0.16) & 49.6 (0.9) & 2.02 (0.07) & 0.58 (0.17) & 50.1 (0.9) & 2.03 (0.07) \\
$10^{-1}$ & 1.01 (0.24) & 48.6 (2.2) & 1.97 (0.13) & 0.72 (0.17) & 49.4 (1.3) & 1.98 (0.10) & 0.55 (0.15) & 49.8 (0.9) & 2.00 (0.07) & 0.51 (0.14) & 50.2 (0.9) & 2.02 (0.07) \\
$1$       & 2.31 (0.14) & 49.9 (1.8) & 2.26 (0.16) & 1.77 (0.15) & 49.9 (1.3) & 2.12 (0.12) & 1.26 (0.12) & 50.1 (0.9) & 2.03 (0.07) & 1.04 (0.13) & 50.4 (0.9) & 2.01 (0.06) \\
\bottomrule
\end{tabular}
\end{minipage}
\end{sideways}

\end{document}